\documentclass[aps,10pt,twocolumn,floats,floatfix,amsmath,amssymb,prb,raggedbottom]{revtex4-2}
\usepackage{graphicx}
\usepackage{color}
\usepackage{xcolor}
\usepackage{bm}
\usepackage{enumerate}
\usepackage{amsmath}
\usepackage{float}
\usepackage{mathtools,tikz}
\usepackage{circledsteps}
\usepackage{pstricks}
\usepackage{oubraces}
\usepackage{nicematrix,tikz}
\usetikzlibrary{tikzmark,calc}
\usepackage[colorlinks]{hyperref}
\usepackage[none]{hyphenat}

\begin{document}
 
\title{Hierarchical analytical approach to universal
spectral correlations \\
in Brownian Quantum Chaos}

\author{Tara Kalsi}
\email{tarakalsi@gmail.com}
\author{Alessandro Romito}
\author{Henning Schomerus}
\affiliation{Department of Physics, Lancaster University, Lancaster, LA1 4YB, United Kingdom}

\begin{abstract}
We develop an analytical approach to the spectral form factor and out-of-time ordered correlators in zero-dimensional Brownian models of quantum chaos. 
The approach expresses these spectral correlations as part of a closed hierarchy of differential equations that can be formulated for all system sizes and in each of the three standard symmetry classes (unitary, orthogonal, and symplectic, as determined by the presence and nature of time-reversal symmetry). 
The hierarchy applies exactly, and in the same form, to Dyson's Brownian motion and all systems with stochastically emerging basis invariance, where the model-dependent information is subsumed in a single dynamical time scale whose explicit form we also establish. 
We further verify this universality numerically for the Brownian Sachdev-Ye-Kitaev model, for which we find perfect agreement with the analytical predictions of the symmetry class determined by the number of fermions. 
This results in a complete analytical description of the spectral correlations and allows us to identify which correlations are universal in a large class of models. 
\end{abstract}

\maketitle

\section{Introduction}
Chaotic quantum systems exhibit a strong degree of universality in their spectral statistics and dynamical correlation functions \cite{Bohigas1984CharacterizationLaws, Stockmann1999QuantumChaos, Haake2018QuantumChaos}.
This universality emerges when the systems are studied on sufficiently small energy scales, over which the Hamiltonian can be modeled by a random matrix drawn from an appropriate 
ensemble chosen according to the symmetries of the system, such as the three Gaussian ensembles (unitary, orthogonal, and symplectic, abbreviated hereafter as GUE, GOE, and GSE)
\cite{Mehta1960OnMatrix, Dyson1962StatisticalI, Dyson1962StatisticalII, Dyson1962StatisticalIII, Dyson1962TheMechanics, Mehta1967RandomLevels, Brody1981Random-matrixFluctuations, Bohigas1984CharacterizationLaws, Stockmann1999QuantumChaos, Haake2018QuantumChaos}. 
Strong arguments for such random-matrix theory (RMT) universality have been formulated for single-particle systems that are disordered or chaotic in the classical limit \cite{Berry1985SemiclassicalRigidity, Beenakker1997Random-matrixTransport, Akkermans2007MesoscopicPhotons, Stockmann1999QuantumChaos, Haake2018QuantumChaos}. 
This has led to precise diagnostics of quantum chaos, among which the spectral form factor (SFF) stands out as a quantity that combines universal signatures of the symmetry class with high sensitivity to system-specific spectral features \cite{Prange1997TheSelf-Averaging, Haake1996SecularMatrices, Haake2018QuantumChaos}.

More recent interest focuses on the analogous picture for many-body systems \cite{Cotler2017ChaosMatrices,Friedman2019SpectralCharge,Kos2018Many-BodyTheory,Gharibyan2018OnsetSystems, Bertini2018ExactChaos, Chan2018SpectralSystems,Dong2025MeasuringProcessors,
Bouverot-Dupuis2024RandomTimes}, where the Sachdev-Ye-Kitaev (SYK) model has emerged as a paradigmatic setting. 
The SYK model is a maximally chaotic quantum mechanical model of $M$ fermions with random all-to-all interactions that is analytically soluble in the large-$M$ limit \cite{Sachdev1993GaplessMagnet, KitaevAlexei1, Polchinski2016TheModel, Bagrets2016SachdevYeKitaevMechanics, Bagrets2017Power-lawModel, Maldacena2016RemarksModel, Rosenhaus2019AnModel,Behrends2020SymmetryModel}.  
The quantum-chaotic nature of the SYK model was first established following the study of the nearest-neighbor level spacing distribution \cite{You2017Sachdev-Ye-KitaevStates}, which adheres to the RMT predictions of the symmetry class  
determined by $M$.
Subsequently, RMT statistics have also been observed in the SFF \cite{Cotler2017BlackMatrices, Saad2019AGravity, Gharibyan2018OnsetSystems, Liu2018SpectralChaos}, as well as in the late-time behavior of out-of-time ordered correlators (OTOCs) which quantify the propensity of a quantum many-body system to scramble information \cite{KitaevAlexei1, Roberts2018OperatorModel, Kitaev2018TheDual, Bagrets2017Power-lawModel}. 
Since quantum black holes are conjectured to be the fastest scramblers in nature \cite{YasuhiroSekino2008FastScramblers, Lashkari2013TowardsConjecture} and the SYK model saturates the Maldacena chaos bound on scrambling dynamics \cite{Maldacena2016AChaos, Maldacena2016RemarksModel}, the SYK model has also been developed as a template of a boundary theory of the two-dimensional (2D) Jackiw-Teitelboim bulk dilaton gravity theory \cite{Blommaert2018ThePerspective, Blommaert2019FineGravity, Rosenhaus2019AnModel, Trunin2021PedagogicalGravity}. 
These characteristics render the SYK model particularly desirable for the study of quantum chaos and scrambling. 

An important variant of these models generalizes their dynamics to acquire a stochastic time dependence. 
This is in the spirit of studying physical systems with some kind of external noise, which is often taken or assumed to be of a Brownian nature. 
Indeed, introducing randomness or stochasticity to a static model typically promotes solvability and enables one to approximate the average statistical properties of the detailed dynamics of complex physical systems, which aligns well with the RMT framework. 
Within RMT, the paradigm of these models is known as Dyson's Brownian motion (DBM). 
Historically defined via the Brownian motion of the eigenvalues of a Hermitian matrix \cite{Dyson1962AMatrix}, DBM is based on a continuous-time stochastic process generated by unitary matrices and hence may also be reformulated as a dynamical process in its own right \cite{Ito1950BrownianGroup, Yosida1952OnSpace, Hunt1956Semi-groupsGroups, Katori2016BesselModel, Erdos2017ATheory, Schomerus2022NoisySystems, Forrester2024Dip-ramp-plateauUN}. 
For DBM generated by Hamiltonians from the GUE, exact results for the instantaneous spectral statistics have been established in the framework of orthogonal polynomials \cite{Onofri1981SUNAction, Andrews1984LatticeFunctions, Forrester2024Dip-ramp-plateauUN}. 
In the many-body setting, the corresponding paradigm is the Brownian SYK (BSYK) model \cite{Saad2019AGravity,Sunderhauf2019QuantumInformation,Agarwal2022EmergentScrambling, Xu2025DynamicsModel}, which captures the fact that realistic interactions are few-body \cite{French1970ValiditySystems, French1971SomeInteractions, Bohigas1971SpacingHamiltonians}. Nonetheless,
the stochastic interactions dynamically drive the system over time into a featureless infinite-temperature state. 
The subsequent stochastic unitary ensembles governing the dynamics are tractable proxies for chaotic quantum systems since they capture the essence of the evolution without relying on system specifics and therefore typically replicate universal features of the chaotic dynamics. 
Stochastic models of this kind are generally known as Brownian models of quantum chaos, and a key question is again to which extent their spectral statistics are universal. 

In this paper, we develop an analytical approach to study the spectral statistics in such Brownian models at instantaneous times and establish their universality when resolved on a single well-defined dynamical time scale that characterizes the gradual evolution towards full RMT ergodicity within a given symmetry class. 
Our approach follows the spirit of the moment method of DBM \cite{Tao2012TopicsTheory} but significantly relaxes its assumptions to only require stochastically emergent basis invariance, which we argue to hold generically in zero-dimensional stochastic models. 
By explicitly evaluating ensemble averages over this universal data in an incremental stochastic time step, we derive, for each symmetry class, a closed hierarchy of differential equations 
for the instantaneous spectral form factor \cite{Kos2021ChaosDriving, Kalsi2024SpectralScrambling, Forrester2024Dip-ramp-plateauUN}, in which the time derivative of the $n$th order can be expressed in terms of instantaneous spectral correlators of order $m\leq n$. 
From this, we obtain exact analytical expressions for the time evolution of the instantaneous spectral form factor and a whole range of other spectral correlators, including those relevant for the ensemble-averaged OTOC, valid for all times and system sizes $N$.  
We also provide the exact form of the universal ergodic time, which can be evaluated analytically both in DBM and in the BSYK model, and verify our results further by comparison with numerical implementations of these models. 
In particular, the numerical data for the BSYK model agree perfectly with the analytical predictions in the symmetry class determined by the number of fermions, which follow the same structure as in the stationary case. 
Finally, based on these findings, we can also identify the general scope of universality within basis-invariant models.

We establish our results in detail as follows. 
Section~\ref{sec:SFF} reviews the spectral form factor and its stroboscopic and instantaneous variants for time-dependent problems as the central quantities of interest, before introducing the specific Brownian stochastic models studied here. With these preparations, we formulate in Section~\ref{sec:program} the precise program of this work. This is then carried out in the subsequent sections, first for DBM and then for a more general class of Brownian models.
We start by deriving the spectral hierarchy of the unitary DBM model in Section~\ref{sec:hierarchy}. This recovers the exact results for this symmetry class from the framework of orthogonal polynomials \cite{Forrester2024Dip-ramp-plateauUN}, extends them to a wider class of correlators, and allows us to determine their uniform asymptotics for large system sizes $N$ at all times and orders. 
In Section~\ref{sec:goegse}, we extend these considerations to the orthogonal and symplectic symmetry classes, where we establish exact and asymptotic results as well as a formal connection between both classes.
In the next two sections, we turn to the question of universality in more general Brownian models.
In Section~\ref{sec:invariance}, we rederive the exact same hierarchies under the significantly relaxed assumption of basis invariance, which we argue to hold generically in zero-dimensional Brownian models. We also establish the general form of the renormalized ergodic time scale in terms of the stochastic Hamiltonians that generate the dynamics.
In Section~\ref{sec:hierarchySYK}, we verify this universality for BSYK dynamics numerically, where we find precise and detailed agreement with the analytical predictions in all three symmetry classes. 
We conclude the paper with an additional application of our framework and a more general discussion of the observed universality.
The application is given in 
Section~\ref{sec:OTOC} and concerns the ensemble-averaged OTOC that, in the studied Brownian models, can be expressed analytically in terms of quantities derived at most from the second level of the spectral hierarchy.
The discussion of the observed universality, Section~\ref{sec:broader}, emphasizes 
the significance of the U(1) invariance of the studied correlators, which is not obeyed, for instance, by the moments of the density of states, or when one truncates the correlators to their connected parts.
Our general conclusions are formulated in Section~\ref{sec:conclusions}, while further details of the derivations are presented in the Appendices.

\section{Setting the scene \label{sec:SFF}}
In this section, we provide some background on the SFF as a probe for dynamically induced correlations between energy levels and introduce its instantaneous version in the setting of the Brownian stochastic processes studied in this work. 

\subsection{SFF as a diagnostic of quantum chaos}
The arguably most universal signature of quantum chaos is the repulsion of adjacent energy levels \cite{Mehta1967RandomLevels, Bohigas1984CharacterizationLaws, Guhr1998Random-matrixConcepts, Stockmann1999QuantumChaos, Haake2018QuantumChaos}.
This level repulsion can be characterized by the distribution $P(s)$ of the nearest-neighbor level spacings $s$, which for small $s$ displays a universal power law dependence, $P(s)\sim s^\beta$, in the three standard Wigner-Dyson symmetry classes ($\beta=1,2,4$ for the orthogonal, unitary, and symplectic classes, respectively). In contrast, $P(s)\sim s^0$ for Poissonian level statistics, which is representative of the behavior of integrable systems.

Unlike the nearest-neighbor level spacing distribution, the SFF is sensitive to eigenvalue correlations on all scales \cite{Stockmann1999QuantumChaos, Haake2018QuantumChaos}. 
The SFF is given by the Fourier transform of the spectral two-point correlation function and, for autonomous systems with an effectively finite-dimensional Hilbert space, can be expressed as 
\begin{equation} \label{eq:FT}
  K(t) = 
  \overline{\mathrm{tr}\,U(t)\,\mathrm{tr}\,U^\dag(t)} = 
  \overline{\sum_{nm} e^{-i(E_n-E_m) t}}, 
\end{equation} 
where the unitary time-evolution operator $U(t)=e^{-iHt}$ generates the dynamics governed by the Hamiltonian $H$ over time $t$ and $E_i$ are the eigenenergies of $H$ 
\cite{Prange1997TheSelf-Averaging, Brezin1997SpectralTheory, Stockmann1999QuantumChaos, Haake2018QuantumChaos, Berry1985SemiclassicalRigidity, Sieber2001CorrelationsStatistics, Muller2004SemiclassicalChaos, Muller2005Periodic-orbitChaos, Richter2022SemiclassicalChaos, Garcia-Garcia2016SpectralModel, Cotler2017BlackMatrices, Cotler2017ChaosMatrices, Gharibyan2018OnsetSystems, 
Chan2018SolutionChaos, Chan2018SpectralSystems, Kos2018Many-BodyTheory, Bertini2018ExactChaos, Friedman2019SpectralCharge, Saad2019AGravity, Yang2020QuantumHamiltonian, Joshi2022ProbingSimulators, Dag2023Many-bodyAtoms, Dong2025MeasuringProcessors,Bouverot-Dupuis2024RandomTimes, Das2024ProposalDetection}. 
While the SFF is sometimes normalized such that $K(0) = 1$, we adopt the common convention for finite-dimensional systems in which $K(t)$ is such that $K(0) = N^2$, where $N$ is the Hilbert space dimension. 
The SFF is not self-averaging \cite{Prange1997TheSelf-Averaging} and consequently requires an additional ensemble average over statistically similar systems, denoted by the overline $\overline{[...]}$.

The SFF is a useful diagnostic of quantum chaos as its specific shape relates the level statistics to the dynamics of the system \cite{Brezin1997SpectralTheory, Prange1997TheSelf-Averaging, Stockmann1999QuantumChaos, Haake2018QuantumChaos}. 
In contrast, the OTOC, which measures a notion of scrambling that is in turn typically associated to chaos, is insufficient to diagnose chaos when employed as a standalone quantity \cite{Dowling2023ScramblingChaos, Zanardi2021InformationSystems, Carolan2024OperatorSystems, Tripathy2024QuantumLattices, Hornedal2022UltimateComplexity, Hashimoto2023}. For the unitary symmetry class describing Hamiltonians with no additional symmetries, the fingerprint of the aforementioned level repulsion is the ramping up of the SFF as
$K(t)\propto t$ until the Heisenberg time $t_H$, at which the discreteness of the spectrum is resolved and $K(t)$ plateaus at the value $N$. 
This ramp is preceded by an initial dip from $K(0)=N^2$ to $K(t_\mathrm{erg})\sim 1$, which describes the establishment of chaotic dynamics over the ergodic time $t_\mathrm{erg}$. The details of this paradigmatic dip-ramp-plateau structure contain information about the symmetry class, while systematic deviations can be used, for instance, to analyze the onset of thermalization in structured systems \cite{Chan2018SolutionChaos, Singh2021SubdiffusionConstraints}. In particular, for Poisson level statistics, the SFF $K(t)=N$ is flat for all times $t>0$.

\begin{figure}[t]
  \centering
  \includegraphics[width=0.9\linewidth]{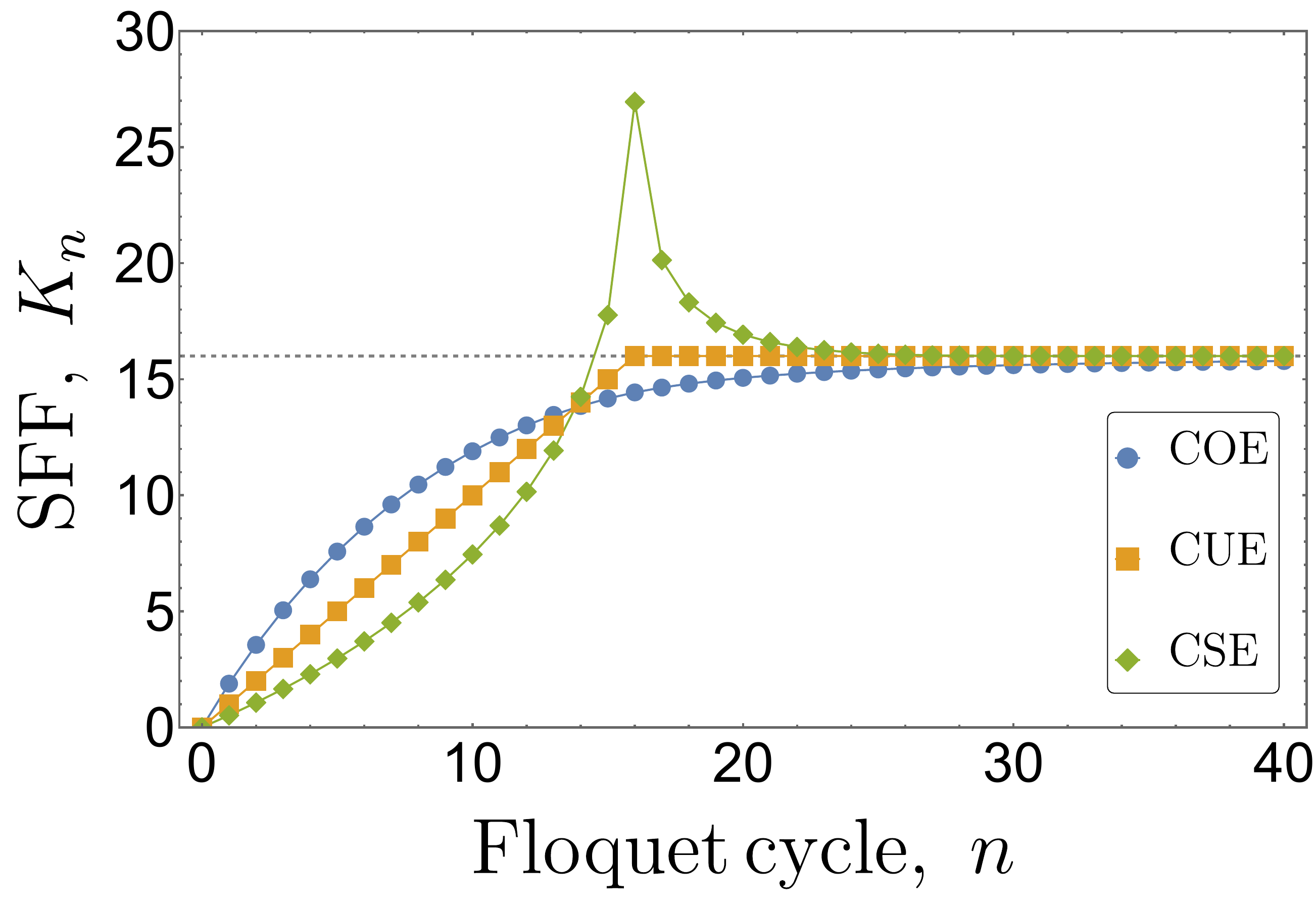}
  \caption{RMT predicts that the stroboscopic
  SFF $K_n$ [Eq.~\eqref{eq:stroboscopicSFF}] of quantum-chaotic systems assumes a paradigmatic (dip)-ramp-plateau structure that takes different forms in different symmetry classes \cite{Haake1996SecularMatrices, Haake2018QuantumChaos}. This is here illustrated for a fixed matrix dimension $N=16$ for the three standard classes (unitary, orthogonal, and symplectic, as based on their corresponding circular ensembles). The dashed line indicates the level of the plateau attained at large $n$, which also coincides with the $n$-independent prediction from Poissonian level statistics in integrable systems. We study how Brownian models of quantum chaos approach the RMT behavior over time $t$, for which we utilize the analogously defined instantaneous SFF $K_n(t)$ [Eq.~\eqref{eq:instantaneousSFF}].
  }
  \label{fig:sff}
\end{figure}

These features take a particularly simple form in a stroboscopic setting, such as that realized by quantum maps or in periodically driven dynamics with period $T$ \cite{Haake2018QuantumChaos}. Using the Floquet operator $U(T)$ so that $U(nT)=U^n(T)$, the natural extension of Eq.~\eqref{eq:FT} gives the stroboscopic SFF
\begin{equation} \label{eq:stroboscopicSFF}
  K_n \equiv \overline{|\, \mathrm{tr} \,U^n(T)|^2}. 
\end{equation}
The theory can then be formulated directly in terms of the unitary operators $U(T)$, which in RMT are drawn from a corresponding circular ensemble (CUE, COE, and CSE).
The CUE applies to fully ergodic models with no additional symmetries, for which RMT then predicts
\begin{equation}
\label{eq:sffcue}
    K_n=N^2\delta_{n0}+\mathrm{min}\,(n,N).
\end{equation} 
This precisely replicates the dip-ramp-plateau behavior with $t_\mathrm{erg} = 1$ and $t_H = N$, with an exactly linear ramp followed by an abrupt onset of a 
perfectly flat plateau.
This behavior is depicted in Fig.~\ref{fig:sff} along with the corresponding results for the other two ensembles, which implement constraints arising from time-reversal symmetry \cite{Haake1996SecularMatrices,Haake2018QuantumChaos, Stockmann1999QuantumChaos}.
These constraints recast the SFF as a curve that plateaus gradually for COE dynamics or as the joining of two regimes by a sharp kink in the case of the CSE. 
In contrast, the Poissonian statistics representative of fully integrable systems again translate to a flat SFF that takes the value of $N$ for all stroboscopic times $n\geq 1$ and hence is completely devoid of the dip-ramp-plateau behavior. 

\subsection{SFF as a diagnostic of emergent ergodicity}

In this work, we study the dynamics in systems with an explicit non-periodic time dependence.
The concept of the spectral form factor can then be applied to analyze the spectral features of the time-evolution operator $U(t)$ at any instance of time. 
This leads to the notion of the instantaneous SFF, which is given by \cite{Kos2021ChaosDriving, Kalsi2024SpectralScrambling, Forrester2024Dip-ramp-plateauUN}
\begin{equation} \label{eq:instantaneousSFF}
  K_n(t) \equiv \overline{|\, \mathrm{tr} \, U^n(t)|^2}. 
  \end{equation}
Here, the continuous time $t$ follows the dynamical evolution of the system, where in particular $K_1(t) = K(t)$ in Eq.~\eqref{eq:FT}. 
Furthermore, the discrete time $n$
systematically resolves the spectral information in $U(t)$ by probing a wider range of correlations in its spectrum. Physically, this can also be interpreted as analyzing the stroboscopic dynamics obtained for a freely imposed period $T=t$, according to Eq.~\eqref{eq:stroboscopicSFF}. 

The instantaneous SFF is particularly useful to study how quantum systems establish ergodic characteristics over time. 
For instance, when they are driven into an infinite-temperature state, the endpoint of the dynamics in $t$ would be expected to conform to the RMT predictions for the stroboscopic SFF shown in Fig.~\ref{fig:sff}. 
Detailed insights into the ergodization dynamics are then obtained by studying the instantaneous SFF for earlier times $t$, where the index $n$ resolves the spectral features on different scales. 
Specifically, the instantaneous SFF analyzes the $n$th moments of $U(t)$ and hence fully probes the dynamics over all energy
and time scales \cite{Kos2021ChaosDriving, Forrester2024Dip-ramp-plateauUN}. 

\subsection{Brownian quantum dynamics}
We will analyze this spectral information in
Brownian processes
that evolve stochastically according to 
\begin{equation} \label{eq:stochastic}
  U(t) \rightarrow U(t+dt) = u(t;dt)U(t), 
\end{equation}
in which the unitary matrix $U(t)$ implementing the dynamics is initialized to the identity, $U(0)=\openone$, and updated incrementally by randomly generated unitary matrices $u(t;dt)\simeq\openone$ over a small time step $dt$. 

We express these incremental unitary time-evolution operators in terms of an instantaneous Hamiltonian $h(t)$. 
In the stochastic setting \eqref{eq:stochastic}, it is convenient to rescale the instantaneous Hamiltonian in such a way that $H(t) = h(t)\sqrt{dt}$, which gives
\begin{align} \label{eq:u}
  u(t;dt) &=\exp{\left(-iH(t)\,\sqrt{dt}\right)}, \quad 0<dt\ll 1, 
\end{align}
or, as a stochastically equivalent but manifestly unitary form that is suitable for numerical implementation, 
\begin{equation} \label{eq:uDBM}
  u(t;dt) = \left(\openone- \frac{iH(t)}{2}\sqrt{dt}\right)\left(\openone+ \frac{iH(t)}{2}\sqrt{dt}\right)^{-1}. 
\end{equation}
This delivers a well-defined stochastic process in the limit $dt\rightarrow 0$ that is of Brownian nature if the instantaneous Hamiltonians $H(t)$ at different time steps are independent of each other. 
We next specify these instantaneous Hamiltonians in detail for the models of interest.

\subsection{Dyson's Brownian Motion (DBM)}
\label{sec:dbm}
Originally, DBM describes the Brownian motion of the eigenvalues of a random matrix that is mathematically modeled using the machinery of the Wiener process \cite{Dyson1962AMatrix, Brouwer1995GeneralizedLeads, Tao2012TopicsTheory, Forrester1998CorrelationsConditions, Forrester2024Dip-ramp-plateauUN}. 
This process is generated by stochastic unitary matrices that update the Hamiltonian and can therefore be defined as a dynamical process in its own right, which then takes the form of a Brownian motion on the group $\mathrm{U}(N)$ of $N\times N$ complex unitary matrices that is initialized by the identity matrix \cite{Ito1950BrownianGroup, Yosida1952OnSpace, Hunt1956Semi-groupsGroups, Katori2016BesselModel, Erdos2017ATheory,Forrester2021DifferentialEnsembles, Forrester2022Highlow-ensembles, 
Schomerus2022NoisySystems, Kalsi2024SpectralScrambling, Forrester2024Dip-ramp-plateauUN}. 
In the stochastic description given above, this dynamical version of DBM is formally implemented by generating the incremental unitary time steps \eqref{eq:u} from instantaneous Hamiltonians $H(t)$ that are drawn from one of the standard Gaussian random-matrix ensembles \cite{Haake2018QuantumChaos,Stockmann1999QuantumChaos}. 
In the GUE, the normally distributed matrix elements satisfy 
\begin{equation} \label{eq:dbmgueelements}
  \overline{H(t)_{lm}}=0 \quad \mathrm{and}\quad \overline{H(t)_{kl}H(t)_{mn}}=N^{-1}\,\kappa\,\delta_{kn}\delta_{lm}, 
\end{equation}
which we scaled such that 
\begin{equation} \label{eq:propto}
  \overline{H(t)}=0 \quad \mathrm{and}\quad \overline{H^2(t)} = \kappa \openone.
\end{equation}
Here, $\kappa$ is a real positive constant that sets the time and energy scales of the dynamics.

Hamiltonians sampled from the GOE and GSE are distributed analogously subject to constraints from time-reversal symmetry. 
In the GOE, $H=H^T$ so that all matrix elements obey $H_{lm}=H_{ml}$ and are real. 
This modifies their two-point correlations to read as
\begin{align} \label{eq:dbmgoeelements}
\overline{H(t)_{kl}H(t)_{mn}}
&=(N+1)^{-1}\,\kappa\,\left(\delta_{kn}\delta_{lm}+\delta_{km}\delta_{ln}\right).
\end{align}
In the GSE, the matrices are self-dual, 
\begin{equation}
\label{eq:symplectic}
  H=(\sigma_y\otimes \openone_{N/2})H^T(\sigma_y\otimes \openone_{N/2})\equiv H^R,
\end{equation}
where $\sigma_r$ with $r=x,y,z$ denotes the Pauli matrices, and the matrix dimension $N$ is necessarily even.
Such matrices are formed out of $(2\times 2)$-dimensional quaternion blocks
\begin{equation}
  H_{lm}=H_{lm}^{(0)}\openone+i\sum_rH_{lm}^{(r)}\sigma_r \quad(l,m=1,2,\ldots N/2),
\end{equation}
where Hermiticity imposes that all parameters are real.
With this constraint, these parameters are independently normally distributed such that 
\begin{align} \label{eq:dbmgseelements}
\overline{H(t)_{kl}H(t)_{mn}}
&=(N-1)^{-1}\,\kappa\,\left(\delta_{kn}\delta_{lm}-\delta_{km}\delta_{ln}\right)\openone_2.
\end{align}
Both here and in Eq.~\eqref{eq:dbmgoeelements}, we have introduced the scale $\kappa$ such that Eq.~\eqref{eq:propto} still holds.

These features are all that we need to formulate our analytical approach to the instantaneous SFF in DBM.

\subsection{The Brownian Sachdev-Ye-Kitaev (BSYK) model} 
\label{sec:bsykdef}

The BSYK model \cite{Sachdev1993GaplessMagnet,KitaevAlexei1} is a quantum mechanical model of $M$ Majorana fermions $\psi_a$ with random $q$-local, all-to-all interactions that depend stochastically on time. 
The Majorana fermions obey the algebra $\{\psi_a,\psi_b \} = \delta_{ab}$ and are defined in a Hilbert space of dimension $\mathrm{dim}\,\mathcal{H} = 2^{M/2}$. 
Specializing to $q=4$, the BSYK Hamiltonian \cite{Saad2019AGravity} is given by
\begin{align}
  \label{eq:hamiltonian}
  H(t)&= \frac{1}{4!} \sum_{a,b,c,d} J_{abcd}(t) \psi_a \psi_b \psi_c \psi_d \nonumber\\
  &= \sum_{a<b<c<d} J_{abcd}(t) \psi_a \psi_b \psi_c \psi_d, 
\end{align} 
where the completely antisymmetric coupling tensor $J_{abcd}(t)$ is drawn independently at each instance
of time $t$. 
Each independent element of the instantaneous coupling tensor is a random real number sampled from a Gaussian distribution with vanishing mean and variance
\begin{equation} \label{eq:variance}
  \overline{J_{abcd}(t)J_{a'b'c'd'}(t')} = \delta_{aa'}\delta_{bb'}\delta_{cc'}\delta_{dd'} \delta(t-t')\frac{J^2(q-1)!}{M^{q-1}}, 
\end{equation}
where here $q=4$, and the overline $\overline{[...]}$ denotes an average over realizations.  
Here, $J$ sets an overall energy scale of the system, which we henceforth set to unity.

The BSYK Hamiltonian \eqref{eq:hamiltonian} reduces to that of the conventional SYK model when the couplings are taken to be independent of time.
This autonomous SYK model is argued to exhibit strongly chaotic dynamics for sufficiently densely connected systems and hence to closely obey RMT predictions 
\cite{Xu2020AHolography, Garcia-Garcia2021SparseDuals, Caceres2022SpectralModels, Orman2024QuantumModel}. 
The fact that, in the BSYK model, the couplings are correlated only at the same instance of time makes this model simpler than the regular SYK model, where in particular the system is expected to relax into an infinite-temperature quasistationary state at long times \cite{Saad2019AGravity},
whose instantaneous statistics should then match the description by a random state vector in a suitable RMT ensemble.
We will examine to which extent RMT universality already holds even before the steady state is reached, where we will relate this to the fast emergence of unitary invariance in the time evolution.

In particular, these features should be manifest in the spectral statistics and dynamical correlations of the system. 
As described above, the structure of the SFF for a given dynamical evolution is expected to be sensitive to the symmetry class of the model. 
The autonomous SYK model is no exception \cite{Cotler2017BlackMatrices}. 
While the SFF exhibits ramp and plateau regions in all three scenarios, its detailed structure differs, with a sharp corner in the unitary class, a smooth curve in the orthogonal class, and a kink in the symplectic class. 
This replicates the key features of the circular ensembles depicted in Figure~\ref{fig:sff}. Analogously, we will establish a very direct correspondence of the instantaneous spectral statistics and dynamical correlations between the BSYK and DBM models generated by the different Gaussian ensembles. 

The specifics of this correspondence will therefore depend on the symmetry class of the model, which here is dictated by the number of Majorana fermions $M$. To describe these, we utilize the fact that, for even $M$, the model may be recast in terms of $M_d=M/2$ Dirac fermions $c_i$, which we also exploit in the numerical implementation of the model. 
The Dirac fermions satisfy the usual algebra
\begin{equation}
  \{c_i,c_j^\dag\}=\delta_{ij}, \quad \{c_i, c_j\}=0, \quad \{c_i^\dag, c_j^\dag\}=0, 
\end{equation}
from which one can define Majorana fermions as
\begin{equation}
  \psi_{2i} = \frac{c_i + c_i^\dag}{\sqrt{2}}, \quad \psi_{2i-1} = \frac{i(c_i - c_i^\dag)}{\sqrt{2}}. 
\end{equation} 
The Hamiltonian \eqref{eq:hamiltonian} preserves the total fermion parity $Q=(-i)^{M_d}\prod_{i=1}^{M} \psi_i$ so that it can be decomposed into blocks corresponding to even and odd values of $Q$. 
Furthermore, the SYK model obeys a particle-hole symmetry \cite{Fu2016NumericalInteractions, Fidkowski2011TopologicalDimension, You2017Sachdev-Ye-KitaevStates}, described by the operator 
\begin{equation} \label{eq:particle-hole}
  P = K \prod_{i=1}^{M_d} \left(c_i^\dag + c_i \right),  
\end{equation}
where $K$ implements complex conjugation. The properties of $P$ impose symmetry constraints on the charge parity sectors of the Hamiltonian, where all three Wigner-Dyson symmetry classes are encountered \cite{You2017Sachdev-Ye-KitaevStates, Garcia-Garcia2016SpectralModel, Meidan2014ScatteringSuperconductors}. 
The particular symmetry class is determined by the value of $M \bmod 8$, where the eightfold periodicity in $M$ conforms with the Atland-Zirnbauer symmetry classification \cite{Altland1997NonstandardStructures, Behrends2019TenfoldModel}.
When $M_d = M/2$ is odd, i.e., for $M \bmod 8 = 2\,\mathrm{or}\,6$, $P$ maps the even and odd charge parity sectors onto each other. 
The two sectors are then degenerate and do not have any antilinear symmetry, so the model falls into the unitary symmetry class. 
For even $M_d = M/2$, $P$ maps each sector to itself and can square to $\pm 1$. 
If $M \bmod 8=4$, $P^2=-1$, which when combined with the antilinearity of $P$ means that $P$ cannot map eigenstates to themselves. 
There is Kramers' double degeneracy of all levels within each individual sector, and the corresponding symmetry class is symplectic.
On the other hand, when $M \bmod 8=0$, $P^2=+1$ and there is no protected degeneracy; the corresponding symmetry class is orthogonal. 
This is summarized as
\[ 
\begin{tabular}{|c|c|c|c|c|c|c|}
 \hline
 $M$ & 8 & 10 & 12 & 14 & 16 & 18 \\
 \hline
 symmetry class & O & U & S & U & O & U \\
\hline
\end{tabular}
\]
and so forth. 

Our numerical implementation of the BSYK model exploits this block structure, in which a unitary transformation permutes the Hamiltonian to assume block form, before then evolving the system in the symmetry-reduced space of the smaller dimension $N=2^{(M/2)}/2=2^{(M/2)-1}$, where the dimension of the full Hilbert space is $\mathrm{dim}\,\mathcal{H}=2N$. 
Further details of this block structure are specified in Appendix \ref{sec:blocks}.

\section{Program of this work}
\label{sec:program}
Our main objective is to develop an approach that facilitates the direct analytic study of the instantaneous SFF \eqref{eq:instantaneousSFF} in the described Brownian models of quantum chaos and allows for their comparison in a unified framework. 
This will furthermore enable us to identify the extent of universality in these and a whole range of related spectral correlators, as well as the role of symmetries both at finite and infinite times.

The key technical 
observation is that the incremental update 
\begin{equation} \label{eq:instantaneousSFFupdate}
  K_n(t+dt) \equiv \overline{|\, \mathrm{tr} \, [U(t+dt)]^n|^2} = \overline{|\, \mathrm{tr} \, [u(t;dt)U(t)]^n|^2}
  \end{equation}
of the instantaneous SFF \eqref{eq:instantaneousSFF} under the stochastic evolution \eqref{eq:stochastic} can be evaluated in the limit that $dt \rightarrow 0$ and, when averaged over the ensemble, depends only on spectral correlators up to the given order $n$.
This culminates in a spectral hierarchy: a closed set of linear first-order differential equations in which expressions for the $n$th order contain only terms up to and including the $n$th level. 
Even though the number of different correlators proliferates with $n$, these equations can be solved algebraically, enabling us to derive exact and asymptotic expressions in all three standard Wigner-Dyson symmetry classes.

Notably, we will arrive at the very same hierarchies in two different settings. The first setting is DBM, which is covered in Section~\ref{sec:hierarchy} for the GUE and in Section~\ref{sec:goegse} for the GOE and GSE. The second setting is models with stochastically emerging basis invariance, which we treat analytically in Section~\ref{sec:invariance}. The applicability of this concept to the BSYK model is discussed in Section~\ref{sec:hierarchySYK}. 

As we show in Section~\ref{sec:OTOC}, the lowest orders of the hierarchy also determine the ensemble-averaged OTOC, which we again determine and validate in all three symmetry classes. 
After developing these results in detail, we will then be able to establish more general implications, in particular, about the extent of the observed universality, in Section~\ref{sec:broader}.

Throughout these sections, we aim to give analytical results for the lowest orders of the hierarchy and illustrate these with the higher orders in figures, as well as through comparison with data from the direct numerical implementation of the underlying Brownian models.

\section{
Spectral hierarchy for DBM with unitary symmetry} \label{sec:hierarchy}

In this section, we first carry out the program in detail for DBM in the GUE, in which the stepwise unitary updates \eqref{eq:u} are generated by instantaneous Hamiltonians $H(t)$ that obey Eq.~\eqref{eq:dbmgueelements}. 
This will extend the exact results known for the SFF in this particular model to a larger range of spectral correlators
and will also serve as a starting point for the generalizations to the other symmetry classes in addition to basis-invariant models.

\subsection{The first-order instantaneous SFF and averaging rules}
We start our considerations with the first-order instantaneous SFF $K_1(t)$, hence, for the index $n=1$, which we will soon see to signify the level of this quantity in the spectral hierarchy.

\begin{widetext}

With the dynamics generated by Eqs.~\eqref{eq:stochastic} and \eqref{eq:u}, the ensemble average 
\eqref{eq:instantaneousSFFupdate}
 over a single step of the stochastic evolution, expanded up to order $dt$, takes the form
\begin{align} \label{eq:sff1update}
  {K_1(t+dt)} 
  & =
  K_1(t) + dt \overline{\,\mathrm{tr}\,\left[H(t)U(t)\right]\,\mathrm{tr}\,\left[H(t)U^\dag(t)\right]} 
  - \frac{dt}{2} \left(\overline{\,\mathrm{tr}\,\left[H^2(t)U(t)\right]\, \mathrm{tr}\,U^\dag(t)} + \overline{\,\mathrm{tr}\,U(t)\,\mathrm{tr}\,\left[H^2(t) U^\dag(t)\right]}\right), 
\end{align}
where we can perform the average over the instantaneous Hamiltonians by applying Eqs.~\eqref{eq:dbmgueelements} and \eqref{eq:propto}.
\end{widetext}

Evaluating these averages joins the two arguments of the traces in the second term according to the rule
\begin{align} \label{eq:jointracegue} \overline{C_{AB}}\equiv\overline{\mathrm{tr}\, \left[{H(t)} {A} \right]\,\mathrm{tr}\,\left[{H(t)} {B}\right]} 
   = \frac{\kappa}{N}
  \mathrm{tr}\, [{A}{B}], 
\end{align}
which applies to any given matrices $A$ and $B$ that are independent of the instantaneous Hamiltonian.
The counterpart of this rule performs the average over instantaneous Hamiltonians within the same trace operation and splits the argument into two,
\begin{align} \label{eq:splittracegue}
\overline{D_{AB}}\equiv\overline{\mathrm{tr}\, \left[{H(t)} {A} {H(t)} {B}\right]} 
  = \frac{\kappa}{N}\mathrm{tr}\, [A]\,\mathrm{tr}\, [B],
\end{align}
which we will need later when evaluating other quantities. Applying the averaging rule \eqref{eq:jointracegue} to Eq.~\eqref{eq:sff1update} gives
\begin{align}
  {K_1(t+dt)} &= K_1(t) + N^{-1}\kappa\,dt\overline{\,\mathrm{tr}\,\left[U(t)U^\dag(t)\right]} -\kappa\,dt\,K_1(t) \nonumber\\
  &= K_1(t) + \kappa\,dt\left(1-K_1(t)\right).
\end{align}
Consequently, in the limit $dt\to 0$, we find the first-order linear differential equation
\begin{equation} \label{eq:K1DE}
  \frac{d}{dt}K_1(t)=\dot{K}_1(t)=\kappa \left(1-K_1(t)\right),
\end{equation}
which for the initial condition $K_1(0) = \overline{|\, \mathrm{tr}\, \openone|^2} = N^2$ is solved by an exponential decay, 
\begin{equation}
  K_1(t) = (N^2-1)e^{-\kappa t}+1. 
  \label{eq:k1}
\end{equation}
We observe that $K_1(t)\to 1$ for large times, in agreement with the RMT prediction \eqref{eq:sffcue} for a fully ergodic system.
The inverse rate $1/\kappa \sim t_\mathrm{erg}$ defines the effective time scale for the dynamics on which this ergodic limit is approached. We will develop a general interpretation of this time scale in Section~\ref{sec:invariance}.

\subsection{The second-order instantaneous SFF and emergence of the hierarchy}
We now turn to the second-order of the instantaneous SFF \eqref{eq:instantaneousSFF}, whose incremental update \eqref{eq:instantaneousSFFupdate} is given by
\begin{align}
  K_2(t+dt) &\equiv \overline{| \mathrm{tr}\, [u(t;dt)U(t)]^2|^2}. 
\end{align}
\begin{widetext}
Expanding this expression again up to order $dt$ 
and averaging over the instantaneous Hamiltonians according to the rules
\eqref{eq:jointracegue}
and \eqref{eq:splittracegue}
gives 
\begin{align} \label{eq:K22}
  K_2(t+dt) &=K_2(t) + \kappa\,dt \left[4-2K_2(t) - N^{-1} \left(\overline{\mathrm{tr}\,[U(t)]^2\, \mathrm{tr}\,U^\dag(t)\,\mathrm{tr}\,U^\dag(t)} + \overline{\mathrm{tr}\,U(t)\,\mathrm{tr}\,U(t) \,\mathrm{tr}\,[U^\dag(t)]^2\,}\right)\right]. 
\end{align}
Here, we encounter two new quantities that signal the start of an emerging hierarchical structure, which furthermore encompasses a more comprehensive set of correlators.
To account for these quantities, we introduce the notation
\begin{equation} \label{eq:define}
  \mathcal{K}_{\{.\},\{.\}}(t)=\mathcal{K}_{pq\dots,rs\dots}(t) = \overline{\left(\mathrm{tr}\,[U(t)]^p \,\mathrm{tr}\,[U(t)]^q \cdots \mathrm{tr}\,[U^\dag(t)]^r \,\mathrm{tr}\,[U^\dag(t)]^s \cdots\right)} ,
\end{equation}
where the sequences of integers specify the number of terms of the form $\mathrm{tr}\,[U(t)]^p$ or $\mathrm{tr}\,[U^\dag(t)]^r$. 
By construction, the respective sums over the tallies $\{\cdot\}$ before and after the comma are equal to the same integer $n$, which enumerates the orders of the hierarchy. 
What was previously expressed as $K_n(t)$ therefore becomes $\mathcal{K}_{n,n}(t)$, while the new quantities appearing in Eq.~\eqref{eq:K22} can now be written as 
\begin{align}
  \mathcal{K}_{11,2}(t) &= \overline{\mathrm{tr}\,U(t)\,\mathrm{tr}\,U(t)\,\mathrm{tr}\,[U^\dag(t)]^2}, \nonumber \\
  \mathcal{K}_{2,11}(t) &= \overline{\mathrm{tr}\,[U(t)]^2\,\mathrm{tr}\,U^\dag(t)\,\mathrm{tr}\,U^\dag(t) }, 
\end{align}
so that Eq.~\eqref{eq:K22} 
becomes
\begin{align} \label{eq:K22new}
  \mathcal{K}_{2,2}(t+dt) &= \mathcal{K}_{2,2}(t) + \kappa\,dt \left[4-2\mathcal{K}_{2,2}(t) - N^{-1} \left(\mathcal{K}_{2,11}(t) + \mathcal{K}_{11,2}(t)\right) \right]. 
\end{align}

The incremental updates of these new quantities $\mathcal{K}_{11,2}(t)$ and $\mathcal{K}_{2,11}(t)$ can again be evaluated using the averaging rules \eqref{eq:jointracegue} and \eqref{eq:splittracegue}, where we also encounter the quantity $\mathcal{K}_{11,11}(t)$. Evaluating its derivative in the same manner 
then yields a closed set of first-order linear differential equations, 
\begin{align} \label{eq:system2}
  \dot{\mathcal{K}}_{1,1}(t) &= \kappa \left[1 - \mathcal{K}_{1,1}(t) \right] ,
  \nonumber\\
  \dot{\mathcal{K}}_{2,2}(t) &= \kappa \left[4 - 2\mathcal{K}_{2,2}(t) -N^{-1}\left(\mathcal{K}_{2,11}(t) + \mathcal{K}_{11,2}(t)\right) \right] ,
  \nonumber\\
  \dot{\mathcal{K}}_{2,11}(t) &= \kappa \left[- 2\mathcal{K}_{2,11}(t) -N^{-1}\left(\mathcal{K}_{11,11}(t) + \mathcal{K}_{2,2}(t) - 4\mathcal{K}_{1,1}(t)\right) \right] ,
  \nonumber\\
  \dot{\mathcal{K}}_{11,2}(t) &= \kappa \left[- 2\mathcal{K}_{11,2}(t) -N^{-1}\left(\mathcal{K}_{11,11}(t) + \mathcal{K}_{2,2}(t) - 4\mathcal{K}_{1,1}(t)\right) \right],
  \nonumber\\
  \dot{\mathcal{K}}_{11,11}(t) &= \kappa \left[- 2\mathcal{K}_{11,11}(t) + 4\mathcal{K}_{1,1}(t) -N^{-1}\left( \mathcal{K}_{2,11}(t) + \mathcal{K}_{11,2}(t)\right) \right]. 
\end{align}
In contrast to the single differential equation \eqref{eq:K1DE} for $n=1$, this system of differential equations includes quantities of the first and second order and is thus hierarchical.

Since realizations in which $U(t)$ is replaced by $U^\dagger(t)$ are statistically equivalent, 
we observe the symmetry $\mathcal{K}_{pq\dots,rs\dots}=\mathcal{K}_{rs\dots,pq\dots}$ of all quantities under the exchange of the left and right indices. 
Accordingly, $\mathcal{K}_{2,11}(t)=\mathcal{K}_{11,2}(t)$ so that the dynamics of these two quantities are effectively described by the same equation. This reduces the cumulative number of equations for $n=1$ and $2$ from 5 to 4.
These equations have to be solved according to 
the dynamical initial condition $U(0)=\openone$, which translates to 
\begin{equation} \label{eq:system2initial}
  \mathcal{K}_{1,1}(0) = N^2, \quad \mathcal{K}_{2,2}(0) = N^2, \quad \mathcal{K}_{11,2}(0)=\mathcal{K}_{2,11}(0)= N^3,\quad \mathrm{and}\quad \mathcal{K}_{11,11}(0) = N^4.
\end{equation}
Here, the power of $N$ corresponds to the number of traces in Eq.~\eqref{eq:define} and hence equals the total number of indices.
As the equations are furthermore linear, they can be solved exactly, from which we obtain 
\begin{equation}
K_2(t)=2+
\frac{1}{4} \left(
(N+1)(N-3) N^2
 e^{\frac{2}{N}\kappa t}+(N-1)(N+3) N^2 e^{-\frac{2}{N}\kappa t}-2 (N^2-1)(N^2-4) \right) e^{-2\kappa t}. 
\end{equation}
While this expression is lengthy, it agrees precisely with the result from the theory of orthogonal polynomials \cite{Onofri1981SUNAction,Andrews1984LatticeFunctions,Forrester2024Dip-ramp-plateauUN} (see Appendix \ref{sec:orthog}).

\subsection{Higher orders of the instantaneous SFF}
It transpires that the principles developed above are all that is required to calculate the instantaneous SFF at all orders and determine a comprehensive set of additional correlators along with this. 
Using the general averaging rules \eqref{eq:jointracegue} and \eqref{eq:splittracegue} in conjunction with the notation \eqref{eq:define}, the incremental updates of these spectral correlators can be expressed conceptually as 
\begin{align}
  \dot{\mathcal{K}}_{pq...,rs...}(t) 
  =
  - \kappa n \mathcal{K}_{pq...,rs...}(t)\,
\underbrace{- \frac{\kappa}{N}\left(+pq \mathcal{K}_{(p+q)..,rs...}(t) - pr\mathcal{K}_{(p-r)q..,s...}(t) + ...\right)}_{\text{from combining traces as in Eq.~\eqref{eq:jointracegue}}} \nonumber \\ 
\underbrace{-\frac{\kappa}{N}\left(\sum_{n=1}^{p-1} \frac{p}{2}\mathcal{K}_{n(p-n)q..,rs...}(t) + ...
+\sum_{n=1}^{r-1} \frac{r}{2}\mathcal{K}_{pq..,n(r-n)s...}(t) + ...\right)\,}_{\text{from splitting traces as in Eq.~\eqref{eq:splittracegue}}}. 
\end{align}
The factors $+pq$ and $-pr$ that appear in the process of combining traces reveal that it is convenient to think of the indices on either side of the comma to be of opposite sign, such that combining indices from the same side yields a positive sign, while combining indices from both sides yields a negative sign. 
Moving all indices to one side of the comma so that $\mathcal{K}_{pq...,rs...}(t)
=\mathcal{K}_{pq...(-r)(-s)...}(t)\equiv\mathcal{K}_{(c_i)_{i=1}^m}(t)$, 
where $m$ is the length of the combined index sequence,
we can then compactly write the spectral hierarchy as
\begin{align}
  \dot{\mathcal{K}}_{(c_i)_{i=1}^m}(t) =
  - \kappa n \mathcal{K}_{(c_i)_{i=1}^m}(t)\,
\underbrace{- \frac{\kappa}{N}\sum_{j<k}c_jc_k\mathcal{K}_{(c_i)_{i\neq j,k} (c_j+c_k)}(t)}_{\text{from combining traces as in Eq.~\eqref{eq:jointracegue}}} 
 \underbrace{-\frac{\kappa}{N}\sum_j \frac{|c_j|}{2}\sum_{k=1}^{|c_j|-1}
\mathcal{K}_{(c_i)_{i\neq j} (k \,\mathrm{sgn}\,c_j )(c_j-k\,\mathrm{sgn}\,c_j ) }(t)\,}_{\text{from splitting traces as in Eq.~\eqref{eq:splittracegue}}}. 
\label{eq:compact}
\end{align}
We see that the terms on the right-hand side never increase the order $n$ of the correlator, while the order reduces whenever two indices $c_j+c_k=0$ annihilate, which results in a constant factor $\mathrm{tr}\,\openone=N$. Furthermore, we only encounter correlators in which the sum of signed indices 
\begin{equation}
\sum_{i=1}^m c_i=0
\label{eq:indexsumrule}
\end{equation}
vanishes, which means that all encountered correlators are $\mathrm{U}(1)$-invariant (we return to the general significance of this feature at the end of the paper, see Section~\ref{sec:broader}). 
The description of this closed hierarchy of differential equations is then completed by the initial conditions
\begin{equation}
\mathcal{K}_{(c_i)_{i=1}^m}(0)=N^m,
\end{equation}
where $m$ is again the combined number of indices.

Applied directly to $n=3$,
we find that we must add the relations 
\begin{align} \label{eq:system3}
  \dot{\mathcal{K}}_{3,3}(t) &= \kappa \left[9 - 3\mathcal{K}_{3,3}(t) -6N^{-1}\mathcal{K}_{3,12}(t) \right],
  \nonumber\\
  \dot{\mathcal{K}}_{3,12}(t) &= \kappa \left[- 3\mathcal{K}_{3,12}(t) -N^{-1}\left(3\mathcal{K}_{12,12}(t) + \mathcal{K}_{3,111}(t)+ 2\mathcal{K}_{3,3}(t) - 6\mathcal{K}_{1,1} - 3\mathcal{K}_{2,2}(t)\right) \right],
  \nonumber\\
  \dot{\mathcal{K}}_{3,111}(t) &= \kappa \left[- 3\mathcal{K}_{3,111}(t)-N^{-1}\left( 3 \mathcal{K}_{3,12}(t) + 3\mathcal{K}_{12,111}(t) - 9\mathcal{K}_{2,11}(t)\right) \right] ,
  \nonumber\\
  \dot{\mathcal{K}}_{12,12}(t) &= \kappa \left[- 3\mathcal{K}_{12,12}(t) + 4\mathcal{K}_{1,1}(t) + \mathcal{K}_{2,2}(t)-2N^{-1}\left( \mathcal{K}_{12,111}(t)+ 2\mathcal{K}_{3,12}(t) - 2\mathcal{K}_{2,11}(t)\right) \right] ,
  \nonumber\\
  \dot{\mathcal{K}}_{12,111}(t) &= \kappa \left[- 3\mathcal{K}_{12,111}(t) + 3\mathcal{K}_{2,11}(t)-N^{-1}\left(\mathcal{K}_{111,111}(t) + 3\mathcal{K}_{12,12}(t) - 6\mathcal{K}_{11,11} + 2\mathcal{K}_{3,111}(t)\right) \right] ,
  \nonumber\\
  \dot{\mathcal{K}}_{111,111}(t) &= \kappa \left[- 3\mathcal{K}_{111,111}(t) + 9\mathcal{K}_{1,1}(t) -6N ^{-1}\mathcal{K}_{12,111}(t) \right]
\end{align}
to the system of equations \eqref{eq:system2} obtained for level $n=2$.
\end{widetext}

The symmetry in the left and right indices reduces what would have been the addition of 10 equations to the 6 given independent equations 
\footnote{More generally, the number of equations generated at the $n$th level is $x(x+1)/2$, where $x$ is the number of integer partitions of $n$, thus yielding an integer sequence $s(n)$ describing the number of partitions of $2n$ that are sum-symmetric (Sequence A086737, OEIS Foundation Inc. (2024), The On-Line Encyclopedia of Integer Sequences, Published electronically at https://oeis.org)}. 
In order to solve the equations at the $n$th level, these equations are then combined cumulatively with all equations at lower levels $m<n$.
The total number of independent differential equations at the $n$th level of this hierarchy proliferates with increasing $n$, as
$1,4,10,25,53,119,239,$ and so on.
However, since the system of equations closes at every level, these equations can be solved exactly for any system size $N$, where the instantaneous SFF $K_n(t)$ again agree precisely with those from the theory of orthogonal polynomials (see Appendix~\ref{sec:orthog}). The other correlators contain concrete dynamical information and complete the description consistently, as we revisit later on in Sections~\ref{sec:OTOC} and \ref{sec:broader}.

\begin{figure}[t]
  \centering
  \includegraphics[width=\linewidth]{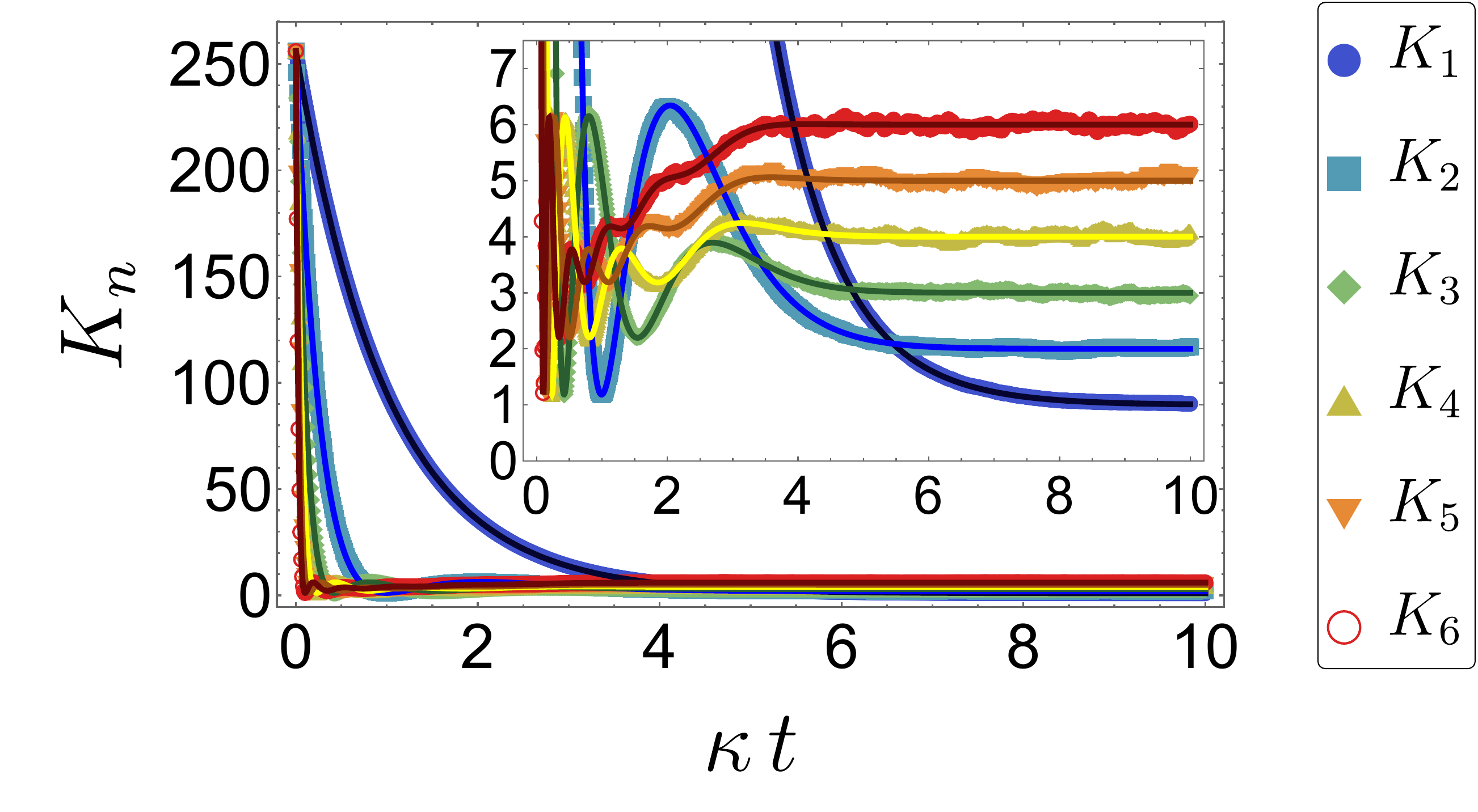}
  \caption{Comparison of the analytical solutions for the instantaneous SFF $K_n(t)$ for unitary DBM, obtained from the spectral hierarchy up to level $n=6$, where $N=16$ (solid colored curves), with numerical sampling of the corresponding dynamics generated by DBM with Hamiltonians drawn from the GUE (data points, obtained from $10^{4}$ realizations with $dt=0.01$). The inset highlights the detailed agreement on intermediate time scales, as well as the expected long-time behavior, where $\lim_{t\to\infty}K_{n}(t)=n$ agrees with the CUE prediction, Eq.~\eqref{eq:sffcue} with $1\leq n\leq N$.}
  \label{fig:DBMtime}
\end{figure}

Figure~\ref{fig:DBMtime} compares the resulting analytical expressions for the instantaneous SFF $K_n(t)=\mathcal{K}_{n,n}(t)$
from the spectral hierarchy up to level $n=6$ with numerical data generated by DBM with instantaneous Hamiltonians drawn from the GUE, where the system size is set to $N=16$. 
The analytic expressions precisely reproduce the non-monotonic structure observed in the numerical sampling of the DBM dynamics over all times and also reproduce the expected long-time behavior that is consistent with the CUE result \eqref{eq:sffcue} for a completely ergodic system. 
As described in Appendix~\ref{app:ergodicother}, this long-time consistency with the CUE also applies to the other correlators appearing in the spectral hierarchy.

\subsection{Uniform asymptotics for large system sizes}
\label{sec:unifasym}

An immediate application of the spectral hierarchy is to determine the asymptotic form of the instantaneous SFF in the limit of large system sizes, $N\to\infty$.
Since, at each order, the SFF dips over short times from $N^2$ to $O(N^0)$ and the large-time limit plateaus at values $O(N^1)$, this requires a systematic uniform expansion in $t$ and $N$. 
This is a more difficult task than obtaining the same asymptotics for the connected part of the correlator \cite{Forrester2024Dip-ramp-plateauUN}, which is by construction only of order $O(N^0)$ but in more general settings involves non-universal contributions from the density of states --- see Section~\ref{sec:broader} for further discussion of these points. 

Using the spectral hierarchy, we achieve this uniform expansion by power counting the terms generated by the averaging rules \eqref{eq:jointracegue} and \eqref{eq:splittracegue}, which then simplify in analogy to the RMT non-crossing approximation (see, e.g., \cite{Brezin1997SpectralTheory}). 
Specifically, all terms that do not maximize the number of traces generated in the averaging rules reduce the power of $N$ by at least two orders, hence a factor of $O(N^{-2})$, unless they are compensated by a trace of the identity matrix. 

In leading order, this asymptotic constraint eliminates all averages from the averaging rule \eqref{eq:jointracegue} unless $A=B^\dagger$. This leads, in the notation of Eq.~\eqref{eq:compact}, to the closed expression
\begin{align}
\mathcal{K}^{(\infty)}_{(c_i)_{i=1}^m}(t)
=\prod_{i=1}^m a_{|c_i|}(\kappa t)
\label{eq:unitaryleading}
\end{align}
of all encountered correlators in their leading order,
where 
\begin{equation}
  a_p(s)=
N e^{-s p/2}\sum_{k=0}^{p-1}\frac{(p-1)!}{(p-k-1)!k!(k+1)!}(-sp)^k.
\label{eq:unitaryleadinga}
\end{equation}
For the instantaneous SFF, this captures the $O(N^2)$ leading-order behavior as 
$K_n(t)\sim a^2_n(\kappa t)$,
where explicitly
\begin{align}
K_1^{(\infty)}&=N^2e^{-\kappa t},\\
K_2^{(\infty)}&=N^2(1-\kappa t)^2e^{-2\kappa t},\\
K_3^{(\infty)}&=N^2\left(1-3\kappa t+\frac{3(\kappa t)^2}{2}\right)^2e^{-3\kappa t}.
\end{align}
This provides a good description of the dip but not of the ramp and plateau.

The desired uniform approximation down to order $O(N^0)$ is obtained by using these leading-order expressions whenever the hierarchy branches according to the averaging rule \eqref{eq:jointracegue}. This automatically guarantees that such branching occurs at most once, which makes the description correct to order $O(N^0)$.
With this simple replacement, the system of differential equations becomes inhomogeneous but closes on each level. From this, we obtain uniform approximations of the form
\begin{align} 
K_1^{(u)}(t)&= K_1^{(0)}, \label{eq:asymptotics1}\\
K_2^{(u)}(t)&=  K_2^{(0)} +\frac{s^4-4 s^3-9s^2}{3}e^{-2s}, \label{eq:asymptotics2}\\
K_3^{(u)}(t)&=  K_3^{(0)} +\frac{3s^2}{8}
\left(9s^4-54 s^3+48s^2-56\right)e^{-3s}, \label{eq:asymptotics3}
\end{align}
where $s=\kappa t$ and
\begin{equation}
  K_n^{(0)}= K_n^{(\infty)}+n(1-e^{-n\kappa t})
  \label{eq:mainterm}
\end{equation}
accounts for the dip and plateau, while the remaining terms provide the detailed oscillations in the ramp.

\begin{figure}[t]
  \centering

  \includegraphics[width=\linewidth]{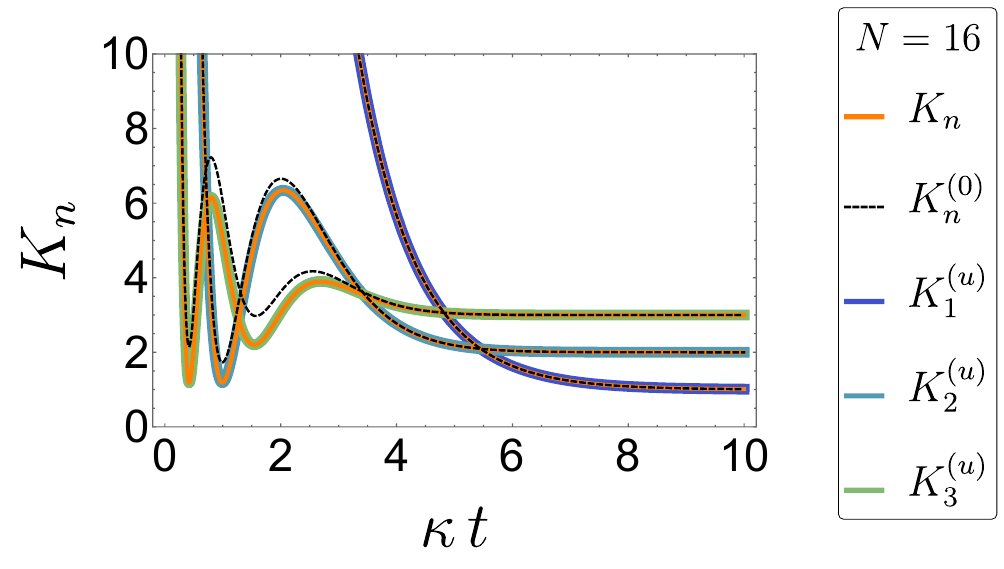}

  \caption{
    Comparison of the exact instantaneous SFF $K_n$ (solid orange curves) with $K_n^{(0)}$ [Eq.~\eqref{eq:mainterm}] (dashed black curves) and the more accurate uniform asymptotic approximations $K_n^{(u)}$ [Eqs.~\eqref{eq:asymptotics1}-\eqref{eq:asymptotics3}] (solid colored curves). 
  }
  \label{fig:asymptotics}
\end{figure}

These different levels of approximations are illustrated in Fig.~\ref{fig:asymptotics}, where they are compared with the exact analytical results for $N=16$. We see that while $K_n^{(0)}$ already captures most of the dynamical behavior, a much more precise agreement is obtained from the uniform approximations $K_n^{(u)}$.

\section{DBM with orthogonal and symplectic symmetry}
\label{sec:goegse}
We now extend the analytical treatment of the instantaneous SFF and related correlators to DBM in the other two standard symmetry classes, where the instantaneous Hamiltonians in Eq.~\eqref{eq:u} are drawn from the GOE or GSE, respectively.
We will maintain the 
composition rule \eqref{eq:stochastic} in which time-reversal symmetry is broken dynamically by the stochastic variation of the Hamiltonian, even though it obeys time-reversal symmetry over each instantaneous time step. A template for this setting would be a system of electrons driven by stochastic electric field variations in the absence of magnetic fields. As we will expand upon further below, the CUE still determines the long-time limit of this dynamics.
Alternatively, we could consider the Brownian motion in one of the corresponding circular ensembles (COE and CSE) itself, where the COE case would be obtained from the unitary dynamics with time-evolution operator denoted as $V(t)$ by setting $U(t)= V^T(t)V(t)$. However, the corresponding correlators are automatically included in the hierarchies of our settings, while the alternative settings physically correspond to the experimentally more challenging situation in which the stochastic signal is itself symmetric in time.

While the resulting hierarchies are much more involved, they remain closed and constitute a complete analytical description for all times $t$, orders $n$, and system sizes $N$. We first develop this for dynamics generated by the GOE and then show that the results for the GSE follow from a simple substitution rule, where one formally replaces the matrix dimension $N$ by $-N$. Furthermore, we will confirm that all hierarchies obey the same long-time limit for finite $N$.

\subsection{Orthogonal symmetry}
\label{sec:DBMGOE}

In the orthogonal symmetry class, the instantaneous SFF updates in a similar fashion to that of the unitary class reported in the previous section but with additional terms and quantities that arise from the symmetric structure of the instantaneous GOE Hamiltonian, whose matrix elements satisfy Eq.~\eqref{eq:dbmgoeelements}.
This modifies the GUE averaging rules \eqref{eq:jointracegue} and \eqref{eq:splittracegue} to read as 
\begin{align} \label{eq:jointracegoe}
  \overline{C_{AB}}
  = \frac{\kappa}{N+1} \left(\mathrm{tr}\, [{A}{B}] + \mathrm{tr}\, [{A}{B^T}] \right) 
\end{align}
and
\begin{align} \label{eq:splittracegoe}
   \overline{D_{AB}}
= \frac{\kappa}{N+1} \left(\mathrm{tr}\,\left[ A \right]\,\mathrm{tr}\,\left[ B \right] + \mathrm{tr}\,[{A} {B^T}]\right). 
\end{align}

Starting from the expansion \eqref{eq:sff1update},
the first-order instantaneous SFF then updates according to
\begin{align}
  {\mathcal{K}_{1,1}(t+dt)} 
  &= \left(1-\kappa\,dt\right)\mathcal{K}_{1,1}(t) 
  + \frac{\kappa\,dt}{N+1}\,\left(N+L(t)\right), 
\label{eq:k11updategoe}
\end{align}
where 
\begin{equation}
  L(t)=\overline{\,\mathrm{tr}\,\left[U(t)U^*(t)\right]}
\end{equation} 
constitutes an additional spectral correlator for level $n=1$ of the hierarchy, and the asterisk denotes complex conjugation. This new quantity incrementally updates---again using the rules \eqref{eq:jointracegoe} and \eqref{eq:splittracegoe} to evaluate the averages---according to 
\begin{align}
  {L(t+dt)}
  &= \overline{\,\mathrm{tr}\,\left[U(t+dt)U^*(t+dt)\right]} \nonumber \\
  &= (1-\kappa\,dt)L(t) +\frac{\kappa\,dt}{N+1}\left(N+ \mathcal{K}_{1,1}(t)\right). 
\end{align}
Altogether, we find the closed system of coupled first-order linear differential equations
\begin{align} 
  \dot{\mathcal{K}}_{1,1}(t)&=\kappa\left((N+1)^{-1} \left(N + L(t)\right) -\mathcal{K}_{1,1}(t)\right), \label{eq:KLDE} \\
  \dot{L}(t)&=\kappa \left((N+1)^{-1}\left(N+ \mathcal{K}_{1,1}(t)\right) -L(t)\right), \label{eq:orthogK1DE}
\end{align}
which for the initial conditions $\mathcal{K}_{1,1}(0) = \overline{|\, \mathrm{tr}\, \openone|^2} = N^2$ and $L(0) = \overline{\, \mathrm{tr}\, \openone} = N$ are solved by 
\begin{align}
\label{eq:k1goe}
  \mathcal{K}_{1,1}(t) &= 1+\frac{N-1}{2} \left(N+2+Ne^{-\frac{2}{N+1} \kappa\, t} \right)e^{-\frac{N}{N+1} \kappa\, t}, \\
  L(t)&= 1+\frac{N-1}{2}\left(N+2-Ne^{-\frac{2}{N+1} \kappa\, t} \right)e^{-\frac{N}{N+1} \kappa\, t}. 
\end{align}

\begin{figure}[t]
  \centering
  \includegraphics[width=\linewidth]{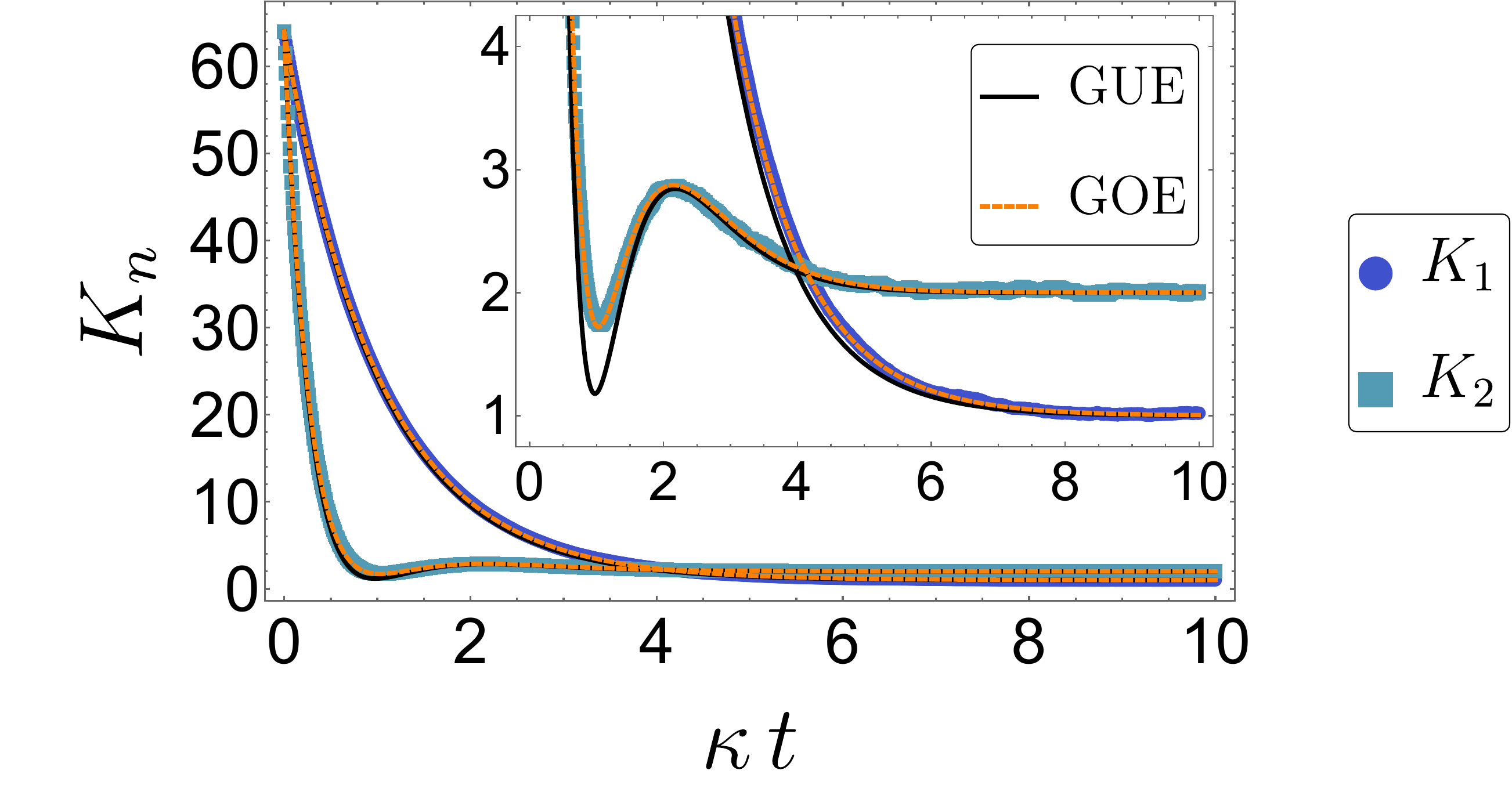}
  \caption{Numerical data for the instantaneous spectral form factor ${K}_{n}(t)$ with $n=1,2$ and $N=8$ for DBM dynamics generated by instantaneous Hamiltonians from the GOE (data points), which match the analytical functions derived for the GOE (dashed curves) and differ clearly from the analytic expressions for the GUE (solid black curves, shown for reference). The DBM data is obtained from $10^4$ realizations with $dt=0.01$.}
  \label{fig:GOEhierarchy}
\end{figure}

When compared to the GUE result, additional spectral correlators therefore arise due to the symmetric structure of the GOE Hamiltonians. Propagating these through the calculations of the higher-order instantaneous SFF means that the number of quantities on each level of the hierarchy proliferates at a much faster rate than in the GUE derivation. Nonetheless, the hierarchy closes at each level and then can be solved analytically, as we expand in more detail for the level $n=2$ in Appendix~\ref{app:goelevel2}, with $K_2(t)$ given by Eq.~\eqref{eq:k22goeresult}.
The analytical results for the instantaneous SFF to levels $n=1$ and $2$ are plotted in Figure~\ref{fig:GOEhierarchy}, where we set the system size $N=8$. There is excellent agreement between the analytical predictions and the numerical data for DBM-generated dynamics with instantaneous Hamiltonians drawn from the GOE, while there exist clear differences when compared to the predictions for GUE dynamics. 

\subsection{Spectral hierarchy in the GSE}

As mentioned above, the results in the symplectic class can be obtained from the results in the orthogonal class by formally replacing the matrix dimension $N$ by $-N$. Here, $N$ refers to the full, even matrix dimension, not the size of the matrix expressed in terms of quaternions. 
From a general perspective, this relation between the two universality classes is inherited from a similar relation between averages in the orthogonal group $\mathrm{O}(N)$ and the compact symplectic group $\mathrm{Sp}(N/2)$ \cite{Collins2006IntegrationGroup}, whose significance we will demonstrate in our second setting, see Section~\ref{sec:invariance}.
In the DBM setting, we can derive this substitution rule concretely by comparing the GOE averaging rules \eqref{eq:jointracegoe} and \eqref{eq:splittracegoe} with their GSE counterparts, which read as
\begin{align} \label{eq:jointracegse}
\overline{C_{AB}}
&= \frac{\kappa}{N-1} \left(\mathrm{tr}\, [{A}{B}] + \mathrm{tr}\, [{A}{B^R}] \right),
\\
\label{eq:splittracegse}
\overline{D_{AB}}
&= \frac{\kappa}{N-1} \left( \mathrm{tr}\,\left[ A \right]\,\mathrm{tr}\,\left[ B \right] - \mathrm{tr}\,[{A} {B^R}] \right).
\end{align} 
Here, $R$ denotes the quaternion conjugation given in Eq.~\eqref{eq:symplectic}, while the traces are the conventional matrix traces, not their quaternion modification.
We now see that the averaging rules for the two ensembles are related by the formal replacements $N\to -N$, $T\to R$, and $\mathrm{tr}\to-\mathrm{tr}$. The same substitution rules then apply to the hierarchy equations and the initial conditions and, therefore, also to the instantaneous SFF itself. 

\begin{figure}[t]
  \centering
  \includegraphics[width=\linewidth]{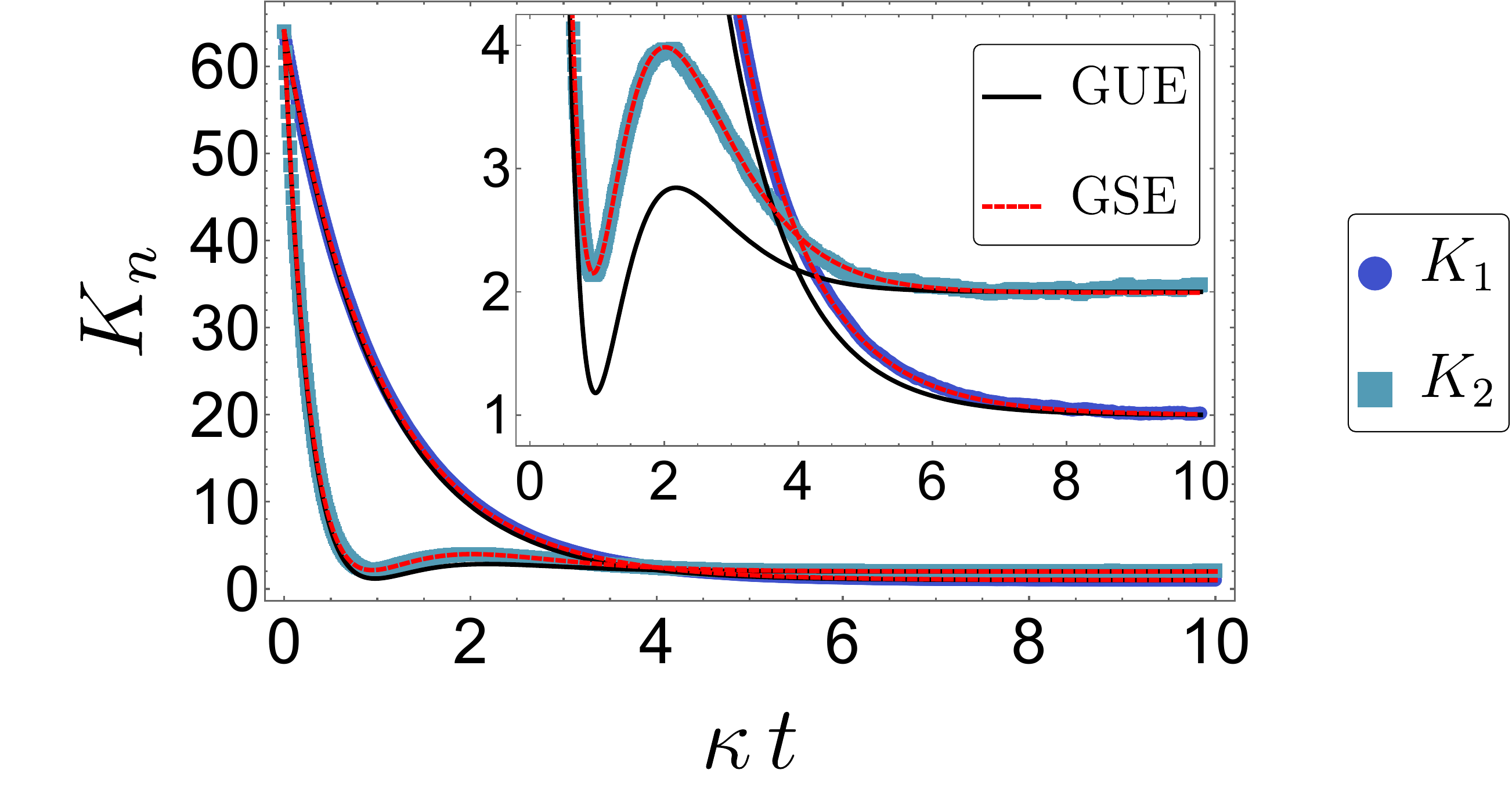}
  \caption{Analogous to Fig.~\ref{fig:GOEhierarchy} but for DBM dynamics generated by instantaneous Hamiltonians from the GSE.}
  \label{fig:GSEhierarchy}
\end{figure}

In this way, we obtain, for instance, from Eq.~\eqref{eq:k1goe} that the first-order instantaneous SFF in the symplectic DBM is given by
\begin{equation}
\label{eq:k1gse}
  K_1(t) = 1+\frac{N+1}{2} \left(N-2+Ne^{\frac{2}{N-1} \kappa\, t} \right)e^{-\frac{N}{N-1} \kappa\, t},
\end{equation}
while the analogous result for $n=2$ is obtained by applying the same substitution to
Eq.~\eqref{eq:k22goeresult}. As shown in Fig.~\ref{fig:GSEhierarchy}, these results again agree perfectly with numerical sampling of the DBM in this symmetry class.

\subsection{Asymptotic relation between the three symmetry classes}
While the analytical expressions of the instantaneous SFF display characteristic signatures in the three ensembles, they also share some common features. These are related to the fact that while the unitary increments $du(t;dt)$ generating the dynamics \eqref{eq:stochastic} obey specific symmetry constraints in the different ensembles, 
the resulting unitary evolution operator $U(t)$ at finite times is not constrained by these symmetries. 
This arises because the unitary increments $du(t;dt)$ in the orthogonal and symplectic symmetry class are confined to cosets, not groups. In the orthogonal class, the coset is formed by transposition-symmetric unitary matrices $du(t;dt)=du^T(t;dt)$, while in the symplectic symmetry class, the coset is that of self-dual unitary matrices $du(t;dt)=du^R(t;dt)$. However, both properties do not hold for products of elements from these cosets.

This results in the convergence of all three ensembles in the limit of long times $t\to \infty$, where all three dynamical versions of DBM converge to the CUE and hence cover all unitary matrices by the Haar measure. This is manifest in the analytical solutions, for instance, where $K_1(t)\to 1$ and $K_2(t)\to 2$ for the given analytical expressions in all three ensembles, and more generally can be verified from the stationary solutions of the spectral hierarchy in all symmetry classes.

It is instructive to compare this convergence for infinite times $t$ with the limit of infinite system sizes $N\to \infty$ at finite times $t$. As in the unitary class (see Section~\ref{sec:unifasym}), we can obtain both a simple leading behavior and more precise uniform asymptotics by applying power counting to the averaging rules.

We start with the leading behavior in the orthogonal class. Power counting then eliminates branching into the term $\mathrm{tr}\,[AB]$ from the averaging rule \eqref{eq:jointracegoe} unless $A=B^\dagger$ and branching into the term $\mathrm{tr}\,[AB^T]$ both in
Eq.~\eqref{eq:jointracegoe} and
Eq.~\eqref{eq:splittracegoe} unless $A=B^*$. The latter combinations can only occur from prior branching of the same form, which disconnects them asymptotically from the leading order of the instantaneous SFF.
Therefore, this leading order coincides with that in the unitary class, see Eq.~\eqref{eq:unitaryleading}.
By application of the stated formal substitution rules, the same argument then also holds for the 
symplectic symmetry class.

Differences between the ensembles appear when we consider the uniform approximation to order $O(N^0)$. The principles of the derivation remain the same as in the unitary class but have to be amended to include the leading orders of the newly generated terms. In the GOE, we obtain 
\begin{align}
  K_1^{(u)}(t)&= K_1^{(0)}(t)+
  s(1+s/2)e^{-s},
  \\
  K_2^{(u)}(t)&=K_2^{(0)}(t)
  -2N(1-s)^2e^{-2s}
  \nonumber\\ 
  &+
  \left[ \vphantom{\frac{1}{3}} \right.\frac{1}{3}(1+s)(3+3s-9s^2+5s^3)\left. \vphantom{\frac{1}{3}} \right]e^{-2s}
  \nonumber\\
  &+2e^{-3s}(N-1)(1-s)+e^{-4s}
  ,
\end{align}
where $K_n^{(0)}(t)$ is given by Eq.~\eqref{eq:mainterm} and $s=\kappa t$.
The results for dynamics generated by the GSE then follow again by formally inverting the sign of $N$.
The asymptotic expression for the first-order spectral form factor agrees between both ensembles, while for larger $n$ the differences are of order $O(N^1)$.

\section{Basis-invariant models}
\label{sec:invariance}

We now show that the same spectral hierarchies can be obtained under a significantly relaxed assumption, which makes them applicable to a much wider class of models with more complex statistical correlations of the Hamiltonian than in the Gaussian ensembles.
For this, we consider Brownian models in which the unitary time-evolution operator $U(t)$ at any time is statistically invariant under a random unitary basis change $V(t)$, such that the transformed operator \begin{equation}
  U'(t)=V(t)U(t)V^\dagger(t)
  \label{eq:basischange}
\end{equation}
occurs with the same probability in the ensemble. 

This assumption is exact in DBM in all three symmetry classes, where $V(t)$ is chosen from the corresponding circular ensemble. More importantly, the same invariance should be well observed in a wide range of Brownian models, as the matrices diagonalizing the unitary generator $du(t;dt)$ are generally far from the identity, even when the generators themselves are arbitrarily close to the identity. Therefore, the eigensystem is expected to become that of an ergodic system on a much shorter time scale than the ergodic time $t_\mathrm{erg}\sim 1/\kappa$ that governs the evolution of the spectral statistics. We will verify this assertion in the next section for the BSYK model.

New forms of the DBM averaging rules
in the different ensembles 
can then be obtained by considering the additional average over $V(t)$. We first work this out for the unitary symmetry class, where the DBM averaging rules are given in \eqref{eq:jointracegue} 
and
\eqref{eq:splittracegue},
and then describe how the same process transfers to the other symmetry classes.

\begin{widetext}
\subsection{Unitary basis invariance}
We note that the basis change \eqref{eq:basischange} also applies in the same form to any power of $U(t)$ and $U^\dagger(t)$. Denoting any such combinations transforming in this way again as $A$ and $B$, momentarily fixing $H$, the average over $V$ from the CUE then gives
\begin{align} \label{eq:jointracecue}
C_{AB}&\equiv\overline{\mathrm{tr}\, \left[{H} V {A} V^\dagger \right]\,\mathrm{tr}\,\left[{H} V{B}V^\dagger \right]} 
  \nonumber\\&= 
  \frac{1}{N^2-1}\left(\mathrm{tr}\,[H^2]-\frac{1}{N}(\mathrm{tr}\,H)^2\right)\mathrm{tr}\, [{A}{B}]
  +
  \frac{1}{N^2-1} \left((\mathrm{tr}\,H)^2-\frac{1}{N}\mathrm{tr}\,[H^2]\right)\mathrm{tr}\, [{A}]\,\mathrm{tr}\, [{B}]
\end{align}
and
\begin{align} \label{eq:splittracecue}
D_{AB}&\equiv\overline{\mathrm{tr}\, \left[{H} V {A} V^\dagger {H} V{B}V^\dagger \right]} 
  \nonumber\\&= 
  \frac{1}{N^2-1} \left(\mathrm{tr}\,[H^2]-\frac{1}{N}(\mathrm{tr}\,H)^2\right)
  \mathrm{tr}\, [{A}]\, \mathrm{tr}\, [{B}]
  +
  \frac{1}{N^2-1} \left((\mathrm{tr}\,H)^2-\frac{1}{N}\mathrm{tr}\,[H^2]\right)
  \mathrm{tr}\, [{A}{B}]
 .
\end{align}
Averaging in the next step also over the Hamiltonian (without any further assumption of its specific ensemble and denoting this additional average again by an overline),
we obtain 
\begin{align} \label{eq:jointracecue2}
  \overline{C_{AB}} &= \frac{1}{N}
  \frac{N^2\mu-\lambda}{N^2-1} \mathrm{tr}\, [{A}{B}]
  +
  \frac{\lambda-\mu}{N^2-1} \mathrm{tr}\, [{A}]\,\mathrm{tr}\, [{B}]
\end{align}
and
\begin{align} \label{eq:splittracecue2}
  \overline{D_{AB}} 
&= 
  \frac{1}{N}
  \frac{N^2\mu-\lambda}{N^2-1}
  \mathrm{tr}\, [{A}]\,\mathrm{tr}\, [{B}]
  +
  \frac{\lambda-\mu}{N^2-1} 
  \mathrm{tr}\, [{A}{B}], 
\end{align}
where all ensemble-specific information is captured by the quantities 
\begin{align}
\lambda&=\overline{(\mathrm{tr}\,H)^2} 
\quad \mathrm{and}\quad
 \mu=\frac{1}{N}\overline{\mathrm{tr}\,[H^2]}.
 \label{eq:lambdamu}
\end{align}
In the GUE, $\lambda=\mu=\kappa$ so that we recover the original DBM averaging rules \eqref{eq:jointracegue} and \eqref{eq:splittracegue}.

We see that in comparison to these original rules, the revised rules contain a first part that only differs by a factor from the original rules and a second part that replicates the trace structure in the initial expression on the left-hand side.
When we revisit the construction of the hierarchy with these revised rules, we see that this results in a minimal modification that amounts to a consistent redefinition of the ergodicity rate $\kappa$ in terms of $\lambda$ and $\mu$.

This modification can be read off directly by comparing the original hierarchy structure \eqref{eq:compact} with the modified structure, which is given by 
\begin{align}
  \dot{\mathcal{K}}_{(c_i)_{i=1}^m}(t) &=
  \underbrace{- \mu n \mathcal{K}_{(c_i)_{i=1}^m}(t)}_{\text{$\kappa$ replaced by $\mu$ }}
\underbrace{- \frac{\lambda-\mu}{N^2-1}\sum_{j<k}c_jc_k\mathcal{K}_{(c_i)_{i=1}^m}(t)}_{\text{retained trace structure in Eq.~\eqref{eq:jointracecue2}}} 
 \underbrace{-\frac{\lambda-\mu}{N^2-1}\sum_j \frac{|c_j|}{2}\sum_{k=1}^{|c_j|-1}
\mathcal{K}_{(c_i)_{i=1}^m}(t)\,}_{\text{retained trace structure in Eq.~\eqref{eq:splittracecue2}}} 
  \nonumber\\
  &
\underbrace{- \frac{\kappa}{N}\sum_{j<k}c_jc_k\mathcal{K}_{(c_i)_{i\neq j,k} (c_j+c_k)}(t)}_{\text{from combining traces in Eq.~\eqref{eq:jointracecue2}}} 
 \underbrace{-\frac{\kappa}{N}\sum_j \frac{|c_j|}{2}\sum_{k=1}^{|c_j|-1}
\mathcal{K}_{(c_i)_{i\neq j} (k \,\mathrm{sgn}\,c_j )(c_j-k\,\mathrm{sgn}\,c_j ) }(t)\,}_{\text{from splitting traces in Eq.~\eqref{eq:splittracecue2}}} 
.
\label{eq:compact2}
\end{align}
\end{widetext}
The two sums in the second line arise from the first part of the modified rules, where the coefficient $\kappa$ now takes the specific form
\begin{equation}
  \kappa=\frac{N^2 \mu-\lambda}{N^2-1} .
 \label{eq:kappanew}
\end{equation}
With this identification, these two sums then already agree with the original hierarchy.
The first line contains two additional sums, which arise from the new second parts of the two modified averaging rules. 
These are all proportional to the same correlator whose time derivative we are determining, which also appears in the very first term now with the coefficient $\mu$.
In these sums, the correlator occurs $n^2$ times multiplied with a positive sign [whenever two oppositely signed indices are combined, which amounts to taking one of $n$ instances of $H$ from a generator $u(t;dt)$ and pairing it with one of $n$ instances of $H$ from a generator $u^\dagger(t;dt)$] and $2\times n(n-1)/2$ times with a negative sign [these are the pairings of $H$ from two generators $u(t;dt)$ or from two generators $u(t;dt)$].
Together with the contribution from the very first term, this combines into a factor
\begin{equation}
-n\mu+n\frac{\lambda-\mu}{N^2-1}=
n\frac{\lambda-\mu+(1-N^2)\mu}{N^2-1}=-n\kappa,
\label{eq:recoverkappa}
\end{equation}
where $\kappa$ is the exact same combination of terms as in Eq.~\eqref{eq:kappanew}.

Altogether, we see that the whole hierarchy remains intact with this specific redefinition of $\kappa$, which holds irrespective of any further details of the ensemble of instantaneous Hamiltonians $H(t)$.
We will return to the discussion of this renormalized ergodic rate after establishing its analogous form in the other two symmetry classes.

\subsection{Orthogonal and symplectic invariance}

Analogous considerations can be carried out in the orthogonal symmetry class, where we impose invariance under orthogonal transformations equipped with the Haar measure on $\mathrm{O}(N)$, and in the symplectic ensemble, where we impose invariance under symplectic transformations equipped with the Haar measure on $\mathrm{Sp}(N/2)$ (where $N$ therefore still refers to the matrix size and hence has to be even; we also keep the conventional notion of a trace for this symmetry class).
\begin{widetext}
 
The derivation can again be carried out on the level of the averaging rules. Carrying out the averages defined in Eqs.~\eqref{eq:jointracecue} and \eqref{eq:splittracecue} with $V$ from the COE, we find
\begin{align} \label{eq:jointracecoe}
C_{AB}&= 
  \frac{N\,\mathrm{tr}\,[H^2] -(\mathrm{tr}\,H)^2}{N(N+2)(N-1)} (\mathrm{tr}\, [{A}{B}]+\mathrm{tr}\, [{A}{B^T}])
  +
  \frac{(N+1)(\mathrm{tr}\,H)^2-2\,\mathrm{tr}\,[H^2]}{N(N+2)(N-1)}\, \mathrm{tr}\, [A]\, \mathrm{tr}\, [B]\,
\end{align}
and
\begin{align} \label{eq:splittracecoe}
D_{AB}&= 
  \frac{N\,\mathrm{tr}\,[H^2]-(\mathrm{tr}\,H)^2}{N(N+2)(N-1)}
  (\mathrm{tr}\,[A]\,\mathrm{tr}\, [B]+
  \mathrm{tr}\, [{A}{B^T}])
  +
  \frac{(N+1)(\mathrm{tr}\,H)^2-2\,\mathrm{tr}\,[H^2]}{N(N+2)(N-1)}
  \,\mathrm{tr}\, [{A}{B}].
\end{align}
Averaging next also over the instantaneous Hamiltonian, 
the rules \eqref{eq:jointracegoe} and \eqref{eq:splittracegoe} for orthogonal DBM therefore become replaced by
\begin{align} \label{eq:jointracecoe2}
\overline{C_{AB}}&= 
  \frac{N^2\mu-\lambda}{N(N+2)(N-1)} (\mathrm{tr}\, [{A}{B}]+\mathrm{tr}\, [{A}{B^T}])
  + \frac{(N+1)\lambda-2N\mu}{N(N+2)(N-1)} \,\mathrm{tr}\, [A]\, \mathrm{tr}\, [B]\,
\\[.2cm]
\label{eq:splittracecoe2}
\overline{D_{AB}}&= 
  \frac{N^2\mu-\lambda}{N(N+2)(N-1)}
  (\mathrm{tr}\,[A]\,\mathrm{tr}\, [B]+
  \mathrm{tr}\, [{A}{B^T}])
  + \frac{(N+1)\lambda-2N\mu}{N(N+2)(N-1)}
  \,\mathrm{tr}\, [{A}{B}], 
\end{align}
where $\lambda$ and $\mu$ remain defined as in Eq.~\eqref{eq:lambdamu}.
\end{widetext}
As in the unitary case, these modified rules contain a part that replicates the original rules \eqref{eq:jointracegoe} and \eqref{eq:splittracegoe} with a modified coefficient $\kappa$, which now takes the form
\begin{equation}
  \kappa=\frac{N+1}{N(N+2)(N-1)}(N^2\mu-\lambda),
\end{equation}
Furthermore, the modified rules contain again an additional part that replicates the trace structure of the terms that have been averaged, which gives rise to the same modification of $\kappa$ when fed into the hierarchy. This follows in analogy to Eq.~\eqref{eq:recoverkappa}, 
where terms now combine according to
\begin{equation}
-n\mu+n\frac{(N+1)\lambda-2N\mu}{N(N+2)(N-1)}=-n\kappa.
\label{eq:recoverkappaagain}
\end{equation}

By an analogous calculation, we have verified that for invariance under symplectic basis changes, the averaging rules again carry over by the replacements $N\to -N$, $\mathrm{tr}\to- \mathrm{tr}$, and $T\to R$.
\subsection{Brief summary of the main result}
Summarizing all three cases, 
we find that basis invariance results in the same spectral hierarchy as that from DBM, where the generalized ergodicity rate can be written as
\begin{align}
  \kappa=\gamma(N^2\mu-\lambda).
  \label{eq:kappaeffall}
\end{align}
Here, $\lambda$ and $\mu$, given in Eq.~\eqref{eq:lambdamu}, are determined by the statistics of the instantaneous
Hamiltonian, while 
\begin{align}
  \gamma&=\frac{1}{(N+1)(N-1)}\quad&\mbox{(unitary class)},
\nonumber  \\
  \gamma&=\frac{N+1}{N(N+2)(N-1)}\quad&\mbox{(orthogonal class)},
\nonumber  \\
  \gamma&=\frac{N-1}{N(N+1)(N-2)}\quad&\mbox{(symplectic class)}.
  \label{eq:gammaall}
\end{align}
We note that in all cases, this coefficient
corresponds to a Weingarten function 
 \cite{Weingarten1978AsymptoticRank, Collins2006IntegrationGroup}
\begin{equation}
\gamma=\overline{|V_{lm}|^2|V_{l'm'}|^2}=\mathrm{Wg}([1][1])
\end{equation}
in the corresponding group, which involves two different matrix elements that are not related by any symmetry.

\subsection{Consistency with DBM}
The assumption of basis invariance holds manifestly in DBM.
To establish consistency between both versions of the hierarchy,
we make use of the detailed definitions of the Gaussian ensembles in Sec.~\eqref{sec:dbm}.
We then have, from the definitions \eqref{eq:lambdamu} and the property \eqref{eq:propto}, the identity $\mu=\kappa$, where $\kappa$ is at this stage to be interpreted as the coefficient defining the Gaussian ensembles. Furthermore, from the
 detailed definitions of these ensembles we also have 
\begin{align}
  \lambda&=\mu\quad&\mbox{(GUE)},
  \nonumber \\
  \lambda&=\frac{2N}{N+1}\mu\quad&\mbox{(GOE)},
\nonumber  \\
  \lambda&=\frac{2N}{N-1}\mu\quad&\mbox{(GSE)}.
\end{align}
Inserting these expressions along with Eq.~\eqref{eq:gammaall} into 
Eq.~\eqref{eq:kappaeffall}, we recover in all cases that the generalized rate coincides with the original coefficient $\kappa$ appearing in the definitions of the Gaussian ensembles.

\section{BSYK dynamics} \label{sec:hierarchySYK}
We now investigate the applicability of the spectral hierarchy to the Brownian SYK model. For this, we first establish the analytical expression of the generalized ergodicity rate and then compare the ensuing predictions for the instantaneous spectral form factor with the numerical implementation of the model, where we cover all three symmetry classes by varying the number of fermions $M$, as described in Section~\ref{sec:bsykdef}.

\subsection{Ergodicity rate $\kappa$}

To obtain the generalized ergodicity rate \eqref{eq:kappaeffall} in the BSYK model, we need to determine the coefficients $\lambda$ and $\mu$ defined in Eq.~\eqref{eq:lambdamu}.
The structure of the BSYK Hamiltonian 
\eqref{eq:hamiltonian} entails that each realization of it is traceless, $\mbox{tr}\,H(t)=0$, so that $\lambda=0$ vanishes identically.
The coefficient $\mu$ then follows from the variance \eqref{eq:variance} of the instantaneous coupling tensor, where we
have to account for the number of its independent elements (defined by the number of ways one can choose 
$q$ from $M$). This gives
\begin{equation} \label{eq:kappa}
  \mu= \frac{1}{q^2}\binom{M}{q} \frac{(q-1)!}{M^{q-1}}, 
\end{equation}
which ties this coefficient to the degree of sparseness in the model. 
The effective ergodicity rate is then given by 
\begin{equation}
\kappa=N^2\gamma \mu,
\end{equation}
where $N=2^{(M/2)-1}$ is the size of a statistically independent block of the Hamiltonian and $\gamma$ is taken from Eq.~\eqref{eq:gammaall} with the symmetry class determined by the number of fermions $M$.

\subsection{Comparison of numerical and analytical results}
 
As mentioned earlier, we will set $q=4$.
In keeping with our desire to test these analytical predictions in the nontrivial regime, we will furthermore focus on models with only moderately large numbers of fermions, which allows us to resolve statistically significant differences between the ensembles.

\begin{figure}[t]
  \centering
  \includegraphics[width=\linewidth]{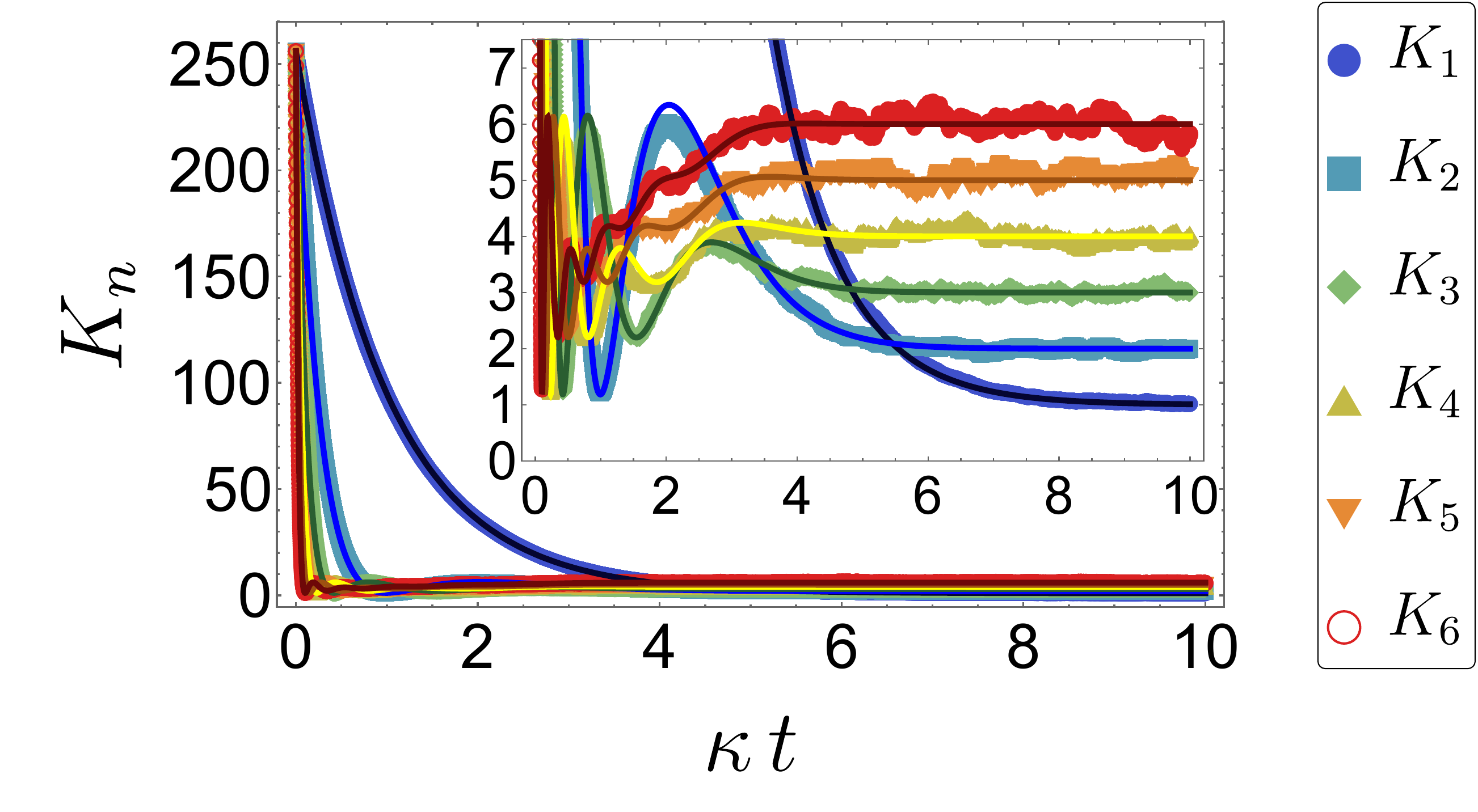}
  \caption{Comparison of the analytical predictions (solid colored curves) for the instantaneous SFF ${K}_{n}(t)$ with data for BSYK-generated dynamics with $M=10$ Majorana fermions, where the instantaneous Hamiltonians belong to the unitary symmetry class. 
  Data are generated from {$10^3$} realizations with $dt=0.01$.}
  \label{fig:BSYKtime}
\end{figure}

Figure~\ref{fig:BSYKtime} shows the numerically obtained time dependence of the instantaneous SFF $K_n(t)$ in analogy to Fig.~\ref{fig:DBMtime} but for numerical dynamics obtained from the BSYK model with $M=10$ Majorana fermions. 
This value of $M$ places the system into the unitary symmetry class, for which the Hamiltonian of dimension $2N=2^5$ can be brought into block-diagonal form, where each block is of dimension $N=16$. As in Fig.~\ref{fig:DBMtime}, we observe excellent agreement with the analytical predictions, with a small deviation only visible in $K_2$.

\begin{figure}[t]
  \centering
  \includegraphics[width=\linewidth]{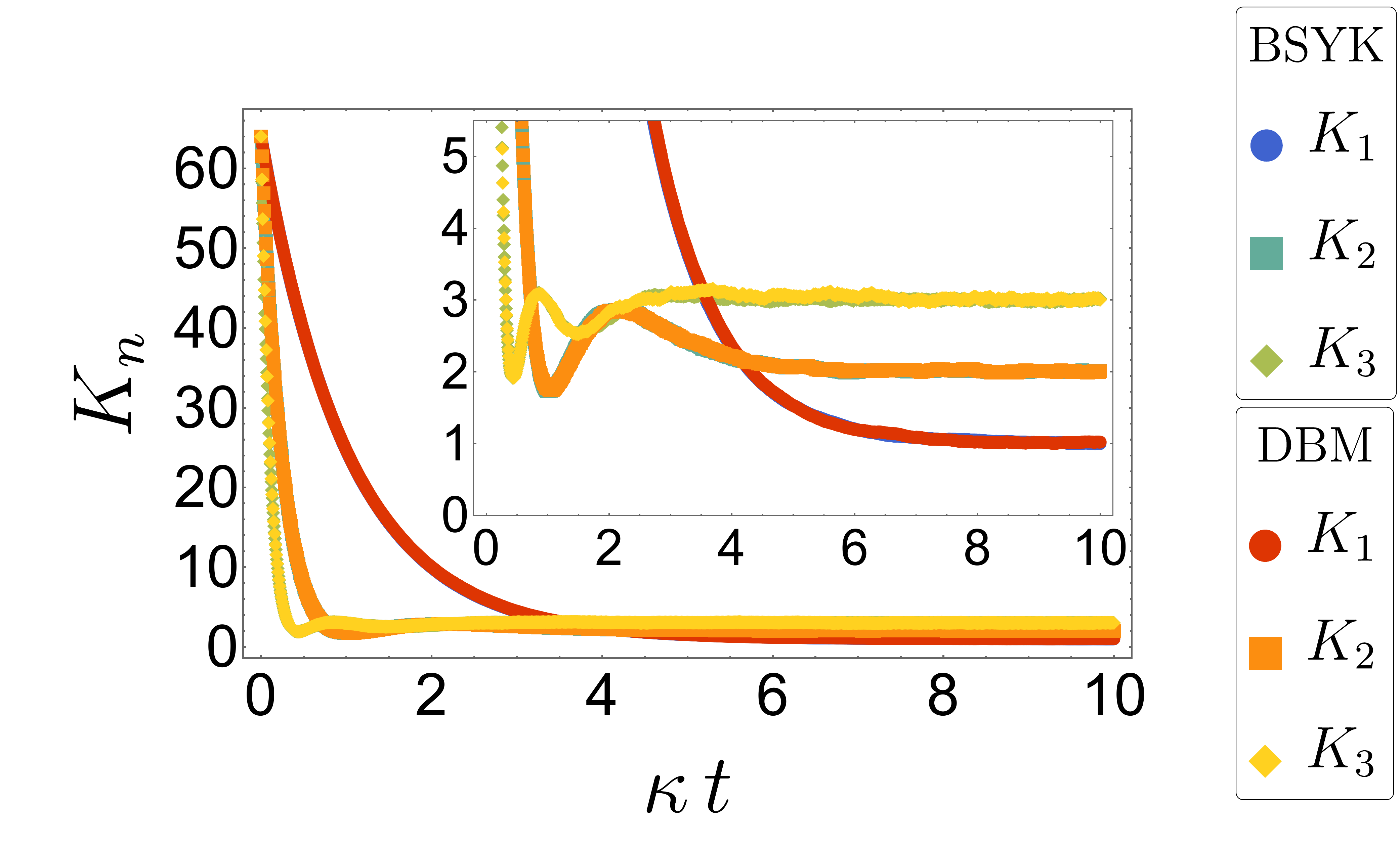}
  \caption{Comparison of numerical data for the instantaneous SFF ${K}_{n}(t)$ for BSYK-generated dynamics of $M=8$ Majorana fermions, where the instantaneous Hamiltonians belong to the orthogonal symmetry class, and DBM-generated dynamics by GOE Hamiltonians of dimension $N=8$. 
  Data are generated for $dt=0.01$ over $10^4$ realizations.}
  \label{fig:goe}
\end{figure}

Figure~\ref{fig:goe} displays numerical data for the BSYK model with $M=8$ Majorana fermions, for which the system is in the orthogonal symmetry class. Instead with the analytical predictions for this case, already shown and contrasted to the unitary case in Fig.~\ref{fig:GOEhierarchy}, we now compare these data with numerical data from GOE DBM-generated dynamics with corresponding matrix dimension $N=8$. We obtain excellent agreement between both Brownian models, which establishes the universality of their spectral correlations now also on the level of the models. We would observe the same agreement if we had plotted the analytical curves instead. 

\begin{figure}[t]
  \centering
  \includegraphics[width=\linewidth]{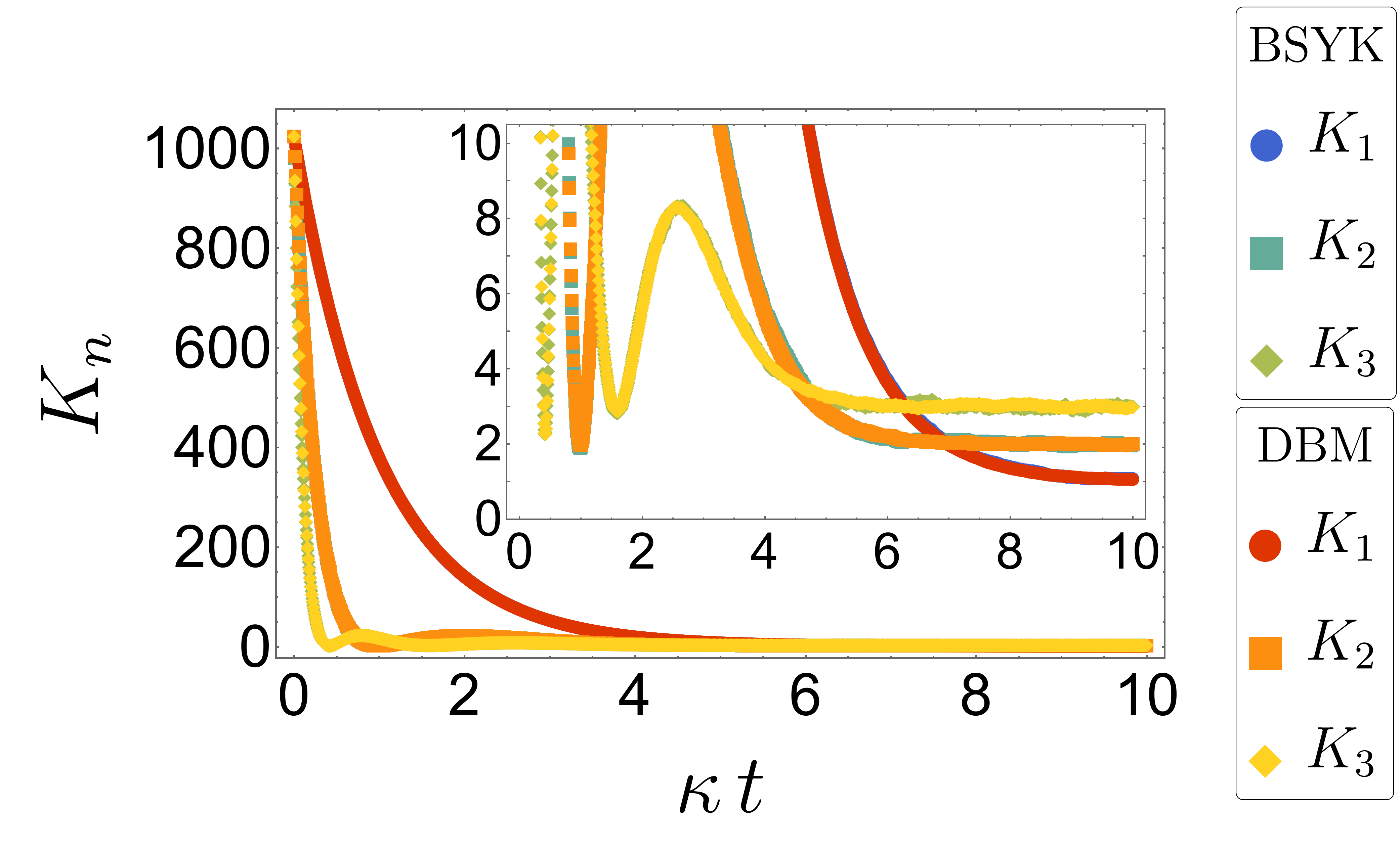}
  \caption{Analogous to Figure~\ref{fig:goe} but for the symplectic class, as realized in BSYK-generated dynamics with $M=12$ Majorana fermions and GSE DBM with $N=32$. 
  {Data are generated for $dt=0.01$ over $10^4$ realizations for DBM and $10^3$ realizations for BSYK.}}
  \label{fig:gse}
\end{figure}

Figure~\ref{fig:gse} is analogous to Figure~\ref{fig:goe} but for the symplectic class, which here is realized for $M = 12$ Majorana fermions ($N=32$). Again, we observe excellent agreement between the BSYK and DBM data, which 
perfectly coincide with the analytical predictions of the spectral hierarchy for this class, see Fig.~\ref{fig:GSEhierarchy}.

Importantly, in all these cases we would have observed noticeable discrepancies if we had equated the ergodicity rate in the BSYK model naively as $\kappa=\mu$. 

\section{Out-of-time-ordered correlator (OTOC)} 
\label{sec:OTOC} 
We now consider the relevance of the correlators appearing in the spectral hierarchy for dynamical correlations. Specifically, we address this in terms of out-of-time-ordered correlators (OTOCs), which serve as a common dynamical
diagnostic of many-body quantum chaos. These correlators originated in the semiclassical description of superconductivity \cite{Larkin1969QuasiclassicalSuperconductivity} but also appear, e.g., as measures of operator growth and information scrambling in quantum many-body systems, including 
random circuit models \cite{Hosur2016ChaosChannels, Roberts2017ChaosDesign, vonKeyserlingk2018OperatorLaws, Khemani2018OperatorLaws, Nahum2018OperatorCircuits, Rakovszky2018DiffusiveConservation,Chan2018SolutionChaos, Mi2021InformationCircuits}, black hole physics \cite{Shenker2014MultipleShocks, Shenker2014BlackEffect, Shenker2015StringyScrambling, Roberts2015LocalizedShocks,
Roberts2016Lieb-RobinsonTheories}, and the SYK model in the large-$M$ limit \cite{KitaevAlexei1, Maldacena2016AChaos, Maldacena2016RemarksModel, Bagrets2017Power-lawModel, Cotler2017ChaosMatrices, Gu2017LocalModels, Roberts2018OperatorModel, Xu2024ScramblingSystems}. 
In the particular setting of quantum-chaotic dynamics, the ballistic behavior at early times allows for the extraction of a Lyapunov exponent, which facilitates the formulation of chaos bounds that have been found to be saturated both in black holes and in the SYK model \cite{Sachdev1993GaplessMagnet, KitaevAlexei1, Maldacena2016AChaos, Maldacena2016RemarksModel}.
We will see that in the Brownian models, ensemble-averaged versions of these correlators can be expressed directly in terms of quantities from the first two levels of the spectral hierarchy. 

\subsection{Definitions}
Per its definition, an OTOC probes the correlation of a time-evolved Heisenberg operator $W(t)$ with an operator $V(0)$ at a fixed earlier time, which is quantified using their squared commutator 
\begin{equation} \label{eq:sqcommutator}
   \mathcal{O}(t) = \langle [W (t),V(0)]^\dag [W (t),V(0)] \rangle.
\end{equation}
In the following, we assume that $V$ and $W$ are observables, thus Hermitian. The name of this correlator arises from the fact that, when expanded, it contains contributions of the form
\begin{equation}
  F(t) = \langle V(0)W(t)V(0)W(t) \rangle, 
\end{equation}
which is of an uncommon structure in terms of the order of time arguments. 

In the stochastic setting \eqref{eq:stochastic}, we consider the OTOC for two generic observables $V(0)=v$ and $W(t)=U^\dag(t) w U(t)$, where the expectation value is evaluated in the infinite-temperature state. Taking the ensemble average, we then build up the OTOC from the quantities
\begin{align} \label{eq:otoc}
 F(t) &= \overline{\mathrm{tr}\,\left[v U^\dag(t) w U(t) v U^\dag(t) w U(t)\right]}
 ,
 \nonumber \\
 G(t) &= \overline{\mathrm{tr}\,\left[v^2 U^\dag(t) w^2 U(t)\right]} ,
 \nonumber \\
 \mathcal{O}(t)&=2\left(G(t)-F(t)\right). 
\end{align}
The OTOC
$\mathcal{O}(t)$ then probes the correlation of the time-evolved operator $w(t)=U^\dag(t) w U(t)$ with the operator $v$ for dynamics starting in an infinite-temperature state.

\subsection{Stationary reference point} 
As a reference point, we first consider the case that $v$ and $w$ are two independent random observables, which for concreteness we model by taking them from the GUE, with normalization $\overline{v^2}=\overline{w^2}=\openone$. The statistical properties of $w$ then transfer unchanged to the time-evolved observable $W(t)$ so that the OTOC is independent of time. This can be made manifest by carrying out the averages explicitly, from which we find
\begin{align}
  \overline{F}&=\overline{\mathrm{tr}\,\left[vwvw\right]} = \frac{1}{N} \nonumber \\ 
  \overline{G}&=\overline{\mathrm{tr}\,\left[v^2 w^2\right]} = N \nonumber \\ 
  \overline{\mathcal{O}}&= 2\left( N - \frac{1}{N}\right). 
  \label{eq:otocstat}
\end{align}
This sets our expectations for the general behavior of the OTOC in the long-time limit, in analogy to the status of the CUE result \eqref{eq:sffcue} for the SFF.

\subsection{OTOC dynamics}

Nontrivial dynamics are observed for the OTOC of a time-evolved operator with itself, which we obtain by equating $v=w$.
Proceeding identically to the construction of the spectral hierarchy in the previous sections, one can consider the incremental updates of the quantities \eqref{eq:otoc} under the stochastic evolution \eqref{eq:stochastic} and derive a system of first-order linear differential equations, which closes under the assumption of basis invariance. This evaluation is simplified by utilizing this basis invariance from the outset, where one again finds convergence with the Gaussian ensembles. 
Implementing this program again for $v$ from the GUE with $\overline{v^2}=\openone$, the ensemble-averaged functions then take the form
\begin{widetext}
\begin{align} \label{eq:otocresult}
  \overline{F(t)} &=\frac{1}{N^2} \left( \,\overline{\mathrm{tr}\,[U(t)]^2 \,(\mathrm{tr}\,U^\dag(t))^2} + \overline{\mathrm{tr}\,[U^\dag(t)]^2\,(\mathrm{tr}\, U(t) )^2 }\,\right) + \frac{1}{N} = \frac{1}{N^2} \left( 2\mathcal{K}_{11,2}(t) +N \right) 
  ,\nonumber \\
  \overline{G(t)} &= N + \frac{1}{N} \left( 1+ \overline{\mathrm{tr}\,U(t)\,\mathrm{tr}\,U^\dag(t)} \right) = N + \frac{1}{N} \left(1 + \mathcal{K}_{1,1}(t) \right)
  ,
  \nonumber \\
  \overline{\mathcal{O}(t)} &= 2N + \frac{2}{N}\mathcal{K}_{1,1}(t) - \frac{4}{N^2}\mathcal{K}_{11,2}(t) = 2 N \left( 1+ \frac{ \mathcal{K}_{1,1}(t)}{N^2} - \frac{2\mathcal{K}_{11,2}(t)}{N^3}\right). 
\end{align}
We see that indeed, as advertised above, the OTOC 
$\overline{\mathcal{O}(t)}$ becomes expressed in terms of quantities from the first two levels of the spectral hierarchy ($n=1,2$). Notably, this also includes the correlator
$\mathcal{K}_{11,2}(t)$. 

Further evaluation depends on the symmetry class of the unitary evolution. 
In the unitary class, where the hierarchy is determined by the differential equations \eqref{eq:system2} with initial conditions \eqref{eq:system2initial}, we find
\begin{align}
\mathcal{K}_{11,2}(t)=\frac{N^2}{4}
\left(
(N-1)(N+3) e^{-\frac{2}{N}\kappa t}-(N+1)(N-3) 
 e^{\frac{2}{N}\kappa t}\right) e^{-2\kappa t},
\end{align}
which delivers the analytical prediction for the OTOC in this symmetry class when combined with Eq.~\eqref{eq:k1} for $\mathcal{K}_{1,1}(t)=K_1(t)$.
The corresponding result for the orthogonal symmetry class is given {by} Eq.~\eqref{eq:k112goe}, while the result for the symplectic symmetry class follows by applying the previously stated substitution rules.
\end{widetext}

In the limit of large $N$, the OTOC turns out to be of order $O(N^{1})$ for all times. This simplifies its asymptotic analysis, for which we only need the leading orders $K^{(\infty)}(t)$ specified in Eqs.~\eqref{eq:unitaryleading} and \eqref{eq:unitaryleadinga}. From this, we obtain the asymptotic prediction
\begin{align}
\overline{\mathcal{O}^{(\infty)}(t)}=2N\left(1+e^{-\kappa t}+2(\kappa t-1) e^{-2\kappa t}\right),
\label{eq:otocinf}
\end{align}
which then applies in the same form to all three symmetry classes.

\begin{figure}[b!]
  \centering
  \includegraphics[width=0.9\linewidth]{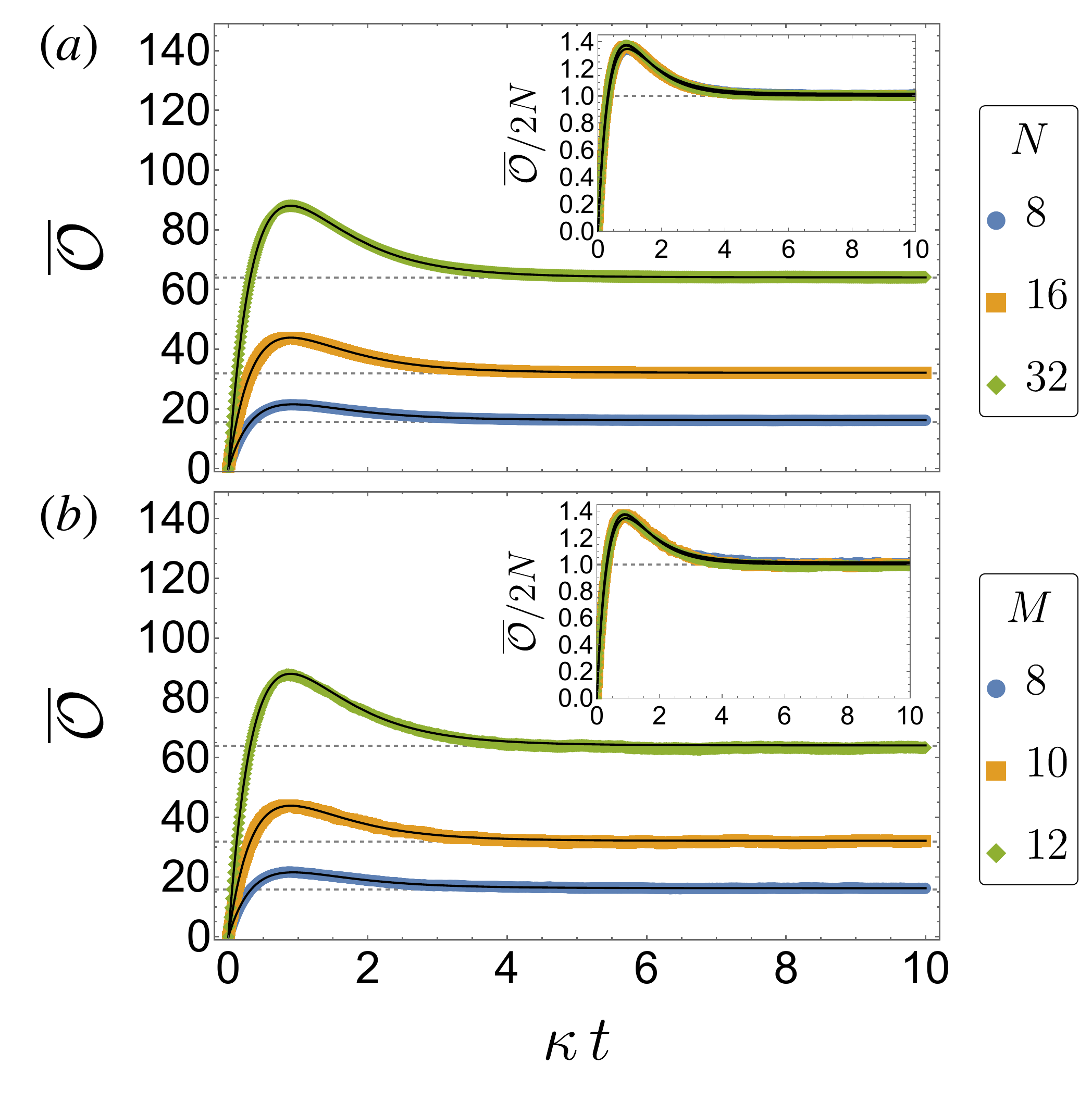}
  \caption{The out-of-time-ordered correlator $\overline{\mathcal{O}(t)}$ calculated numerically for (a) DBM dynamics generated by Hamiltonians sampled from the GOE with dimension $N=8$, the GUE with dimension $N=16$, and the GSE with dimension $N=32$, as well as (b) BSYK dynamics generated by $M=8,10,$ and $12$ Majorana fermions, whose Hamiltonians can be decomposed into two blocks each of dimension $N=8,16,$ and $32$ (blue circles, orange squares, and green diamonds, respectively), plotted as a function of rescaled time, where $\kappa$ is given by Eq.~\eqref{eq:kappa}. This is contrasted with the analytic solutions \eqref{eq:otocresult} derived in the setting of DBM (solid black curves). The inset demonstrates data collapse when the OTOC is rescaled by $2N$, which corresponds to its asymptotic value for large $N$ and $t$. Data are generated over $10^4$ and $10^3$ realizations, respectively ($10^2$ for BSYK-generated dynamics for $M=10$ and $12$).}
  \label{fig:otoc}
\end{figure}

Figure~\ref{fig:otoc} depicts the time behavior of the ensemble-averaged OTOC $\overline{\mathcal{O}(t)}$ in (a) DBM- and (b) BSYK-generated dynamics for varying system dimensions, covering again the three different symmetry classes. With time scaled according to the analytically determined ergodicity rate $\kappa$, the numerical data agree perfectly well both between the two models and with the analytical predictions from Eq.~\eqref{eq:otocresult}. 
In all cases, the OTOC tends over time to the stationary value $\overline{\mathcal{O}}$ given in Eq.~\eqref{eq:otocstat}, which is indicated by the dashed lines.
The respective insets show this numerical data normalized by $2N$. The resulting data collapse demonstrates that even though the chosen matrix dimensions are relatively small, the results for the different symmetry classes {closely follow} the asymptotic prediction \eqref{eq:otocinf}, where all differences between the ensembles fall within statistical bounds.

\section{The broader picture}\label{sec:broader}
The previous section highlights the significance of the entirety of spectral correlators encountered in the framework of this work, beyond those that define the instantaneous SFF itself. To conclude these considerations, we adopt a  broader perspective and set out to delineate the boundaries of the encountered universality in these spectral correlations. 
We will argue that these boundaries themselves align closely with 
all spectral correlators that form the spectral hierarchy.
Our argument rests on the role of time-local U(1) invariance in the Brownian models. As emphasized earlier, all correlators {within} the hierarchy obey this invariance. In the unitary case, this invariance is manifest in the index sum rule \eqref{eq:indexsumrule}, while more generally it follows from the averaging rules in all classes. 

To deepen this argument, we note that this invariance is also manifest in the universal form \eqref{eq:kappaeffall} of the ergodicity rate $\kappa$. In this expression, we can recast the ensemble-specific data as 
\begin{equation}
N^2\mu-\lambda=N\,
\overline{\mathrm{tr}\,[H_0^2]}\equiv N^2\mu_0,
\end{equation}
where $H_0=H-\frac{\mathrm{tr}\,H}{N}\,\openone$ is the traceless part of the Hamiltonian. Therefore, all ensembles in which the Hamiltonians are amended by arbitrarily time-dependent additional contributions $\varphi(t)\openone$ are equivalent, where this holds even when these additions are correlated in time. In the dynamics, these additional contributions then again integrate into an arbitrarily correlated U(1) phase, which however drops out of all of the correlators inside the hierarchy.

We illustrate the relevance of this observation by identifying two closely related, nonuniversal properties. The first property is the density of states of the instantaneous eigenvalues of the time-evolution operator on the unit circle, which can be analyzed in terms of its moments
\begin{equation}
A_n=\overline {\mbox{tr}\,U^n(t)}.
\end{equation}
We see that these moments formally correspond to unbalanced correlators $\mathcal{K}_{n,\cdot}(t)$, in which only one of the two sets of indices appears. 
These moments are not manifestly invariant under arbitrary U(1) transformations. Indeed, they can all be made to vanish by averaging over the global phase, which results in a flat density of states on the unit circle.
In a Brownian model, this can be achieved on a freely chosen time scale by amending the Hamiltonians as mentioned above, while leaving all correlators within the hierarchy unchanged.

The second closely related property is the connected part of the SFF, given by
\begin{equation}
\label{eq:connectedsff}
S_n(t)=K_n(t)-A_n^2(t).
\end{equation}  
This connected part has been the focus of previous studies of unitary DBM, where it is more directly amendable to an analytical treatment as it removes the contributions of order $N^2$. However, utilizing this connected part only, we would not be able to establish the observed universality of spectral correlations beyond DBM as it then mixes in the nonuniversal information from the density of states. Indeed, we see immediately from its definition \eqref{eq:connectedsff} that the connected part does not obey statistical U(1) invariance in the context of these more general models. 

A concrete manifestation can be established directly by inspecting $A_1$, which in our models obeys the differential equation
\begin{equation}
  \dot{A_1}=-\frac{\mu}{2}A_1(t), 
\end{equation}
with initial condition $A_1(t)=N$. This is solved by $A_1(t)=N e^{-\mu t/2}$ with $\mu$ defined as in Eq.~
\eqref{eq:lambdamu},
which therefore serves as an additional independent rate to describe statistical quantities outside the spectral hierarchy.

On the other hand, the leading order asymptotics of these moments in DBM also appear in our analysis, see Eq.~\eqref{eq:unitaryleading}.
This suggests that a larger degree of universality, also encompassing these quantities, could possibly be reinstated by a suitable unfolding of the spectrum (which we expressly did not require for the quantities captured in our framework). Furthermore, one may consider U(1)-invariant versions of these moments, which could be based on the traceless parts of the instantaneous Hamiltonians \footnote{Such moments should be connected with the analysis of DBM on $\mathrm{SU}(N)$, see \cite{Levy2010CentralGroup}.}. 

Finally, we recall that the universality of properties within the hierarchy still rests on an assumption, namely that of rapidly established basis invariance. We argued this to be observed for a wide class of models based on the fact that the matrices diagonalizing the instantaneous unitary operators $du(t;dt)$ are generally far away from the identity. Furthermore, it follows that under the composition of two time steps, the new set of eigenstates is generally a nonperturbative combination of the eigenstates from each step. This stochastically emergent basis invariance is manifest in DBM and  
numerically well observed in the BSYK model.
However, it is of course easy to come up with structured Brownian models that develop basis invariance on scales that compete with the ergodic time, with (non-zero-dimensional) Brownian circuits being a prime example of such models in the literature \cite{Gharibyan2018OnsetSystems,Chan2018SpectralSystems}. An open question following on from our work, therefore, is to identify the exact range of Brownian models that develop basis invariance on quasi-instantaneous time scales.

\section{Conclusions and Outlook} \label{sec:conclusions}

In summary, we described an analytical approach to obtain exact expressions for correlators that describe the instantaneous spectral statistics in zero-dimensional Brownian models of quantum chaos. 
The approach organizes these correlators into a hierarchy, which includes the instantaneous SFF as well as other correlators that feature, e.g., in ensemble-averaged OTOCs.
We established the universality of these spectral correlations encompassing all systems that rapidly develop statistical basis invariance, where we have carried this out in each of the three standard Wigner-Dyson symmetry classes. 
This property applies exactly to Dyson Brownian motion and agrees excellently with numerical data in the Brownian SYK model, in which the symmetry class depends on the 
number of fermions. 

To develop this picture, we presented a specific derivation of the spectral hierarchy for DBM,
and a general derivation that exploits stochastically emergent basis invariance according to the symmetry class in more generic Brownian models. 
The latter scenario results in exactly the same hierarchy with a suitably renormalized ergodic time, whose general form we established in terms of the ensemble of instantaneous Hamiltonians, while its specific form can be determined analytically in the Brownian SYK model. 
To illustrate the analytical results, we furthermore compared them with data from numerical implementations of the stochastic dynamics in these models. 

We also explored the broader significance of these findings.
By examining the assumptions of our derivations, we clarify that the observed universality is confined to correlators that obey statistical U(1) invariance. 
This is manifest in the OTOCs and the full instantaneous SFF but not, for instance, when one investigates  
only its connected part. 
While this makes determining the statistics for large system sizes $N$ in principle more challenging, our approach allows us to establish these asymptotics uniformly in time and $N$.

Overall, our results, therefore, delineate the boundaries of universal spectral statistics over a large range of Brownian models of quantum chaos and make them amenable, in all three standard Wigner-Dyson symmetry classes, to a systematic analytical treatment.
We also envisage that this approach seamlessly carries over to other symmetry classes but leave this open for future work. 
At the present stage, these results can already serve as a useful benchmark to observe deviations from universality in more structured or constrained models. At the same time, it would be interesting to see if the techniques developed here can be extended to account for specific sources of non-universality, such as
finite noise correlation times, non-Gaussian fluctuations beyond the Wiener process, and, ultimately, static backgrounds. This could then serve as a bridge to other paradigmatic settings of many-body quantum chaos, such as those encountered, e.g., in the context of quantum gravity models.

\begin{acknowledgments}
This research was funded by EPSRC via Grant No. EP/T518037/1.
All relevant data present in this publication can be accessed at \cite{data}.
\end{acknowledgments}

%

\newpage

\appendix

\section{Detailed block structure in the BSYK model}
\label{sec:blocks}
As described in Section~\ref{sec:bsykdef}
the BSYK Hamiltonian \eqref{eq:hamiltonian} can be brought into a block form 
\begin{equation}
\begin{pmatrix}
  \cdot & 0 \\
  0 & \cdot
\end{pmatrix}
,
\end{equation}
where each block corresponds to even or odd values of $Q$, and the specific RMT statistics of the blocks are dependent on the action of the particle-hole symmetry operator \eqref{eq:particle-hole} \cite{You2017Sachdev-Ye-KitaevStates, Garcia-Garcia2016SpectralModel}. 
 
Our numerical implementation of Brownian SYK dynamics exploits this block structure: once the Hamiltonian of dimension $2N$ has been reduced to block form, we need only run dynamics for the two blocks each of dimension $N$. 

This is then further simplified by constraints and correlations observed for these blocks in the different symmetry classes.
In the unitary symmetry class, the blocks are formed by a matrix $A$ and its transpose $A^T$, 
\begin{equation}
\begin{pmatrix}
  A & 0 \\
  0 & A^T
\end{pmatrix}. 
\end{equation}
The dynamics obtained from both blocks are spectrally degenerate so that we only need to generate dynamics for one of them.
This degeneracy does not occur in the orthogonal and symplectic symmetry classes, where one instead obtains two independent blocks 
\begin{equation}
\begin{pmatrix}
  A & 0 \\
  0 & B
\end{pmatrix}. 
\end{equation}
In the orthogonal class, both matrices can be brought into symmetric form, 
$A=A^T$ and $B=B^T$, while in the symplectic class they can be brought into self-dual form, $A=A^R$ and $B=B^R$, so that in the latter case they display Kramers' degeneracy.
 
\begin{widetext}

\section{Second-order instantaneous SFF in GOE and GSE DBM}
\label{app:goelevel2}
In DBM generated by the GOE or GSE, the number of correlators at each level of the hierarchy proliferates rapidly due to the additional terms in the averaging rules \eqref{eq:jointracegoe} and \eqref{eq:splittracegoe}. To illustrate the complexity but eventual feasibility of the calculations, we here work through the construction of the second level of the hierarchy in the GOE, in a similar fashion to the first level in Section~\ref{sec:DBMGOE} of the main text.

For reference, let us summarize the first level of the hierarchy in the GOE concisely as
\begin{align}
  a&\equiv K_1(t)=\mathcal{K}_{1,1}(t)=\overline{\mathrm{tr}\,U(t)\,\mathrm{tr}\,U^\dag(t)}, \\
  b&\equiv L(t)=\overline{\mathrm{tr}\,\left[U(t)U^*(t)\right]},
\end{align}
 such that Eqs.~\eqref{eq:KLDE} and \eqref{eq:orthogK1DE}  reduce to
\begin{align}
  \dot{a}&=\kappa\,\left(\nu \left(N + b\right) -a\right), \nonumber \\
  \dot{b}&=\kappa \left(\nu\left( a + N\right) -b\right).
\end{align}
Here the time dependence of $a$ and $b$ is implied and
we have introduced $\nu=(N+1)^{-1}$ for brevity. The initial conditions $a(0)=N^2$ and $b(0)=N$ then determine the solutions
\begin{align}
   a&= 1+\frac{N-1}{2} \left(N+2+Ne^{-2\nu \kappa\, t} \right)e^{-N \nu \kappa\, t}, \\
  b&= 1+\frac{N-1}{2}\left(N+2-Ne^{-2\nu \kappa\, t} \right)e^{-N \nu \kappa\, t}. 
\end{align}

For the second level, we anchor the hierarchy again at the instantaneous SFF, now denoted by 
\begin{equation}
  c \equiv K_2(t)=\mathcal{K}_{2,2}(t). 
\end{equation}
Using the GOE averaging rules, this quantity evolves as
\begin{align}
  \dot{K}_2&= \kappa\left(- 2\underbrace{K_2(t)}_{=c} - 2(N+1)^{-1}\left(\underbrace{\mathcal{K}_{11,2}(t)}_{=d} + \underbrace{\overline{\mathrm{tr}\,{[U(t)U^T(t)]}\,\mathrm{tr}\,[U^\dag(t)]^2}}_{=e} \right) +4(N+1)^{-1}\left(N +\underbrace{\overline{\mathrm{tr}\,\left[{U(t)U(t)}{U^*(t)U^*(t)} \right]}}_{=f} \right)\right), 
\end{align}
or more compactly
\begin{align}
  \dot{c}&= \kappa\left(- 2\,c - 2\nu\left(d + e \right) +\,4\,\nu\,\left(N +
  f\right)\right), 
\end{align}
where we also encounter
\begin{align}
  d&\equiv\mathcal{K}_{11,2}(t)=\overline{\mathrm{tr}\,U(t)\,\mathrm{tr}\,U(t)\,\mathrm{tr}\,[U^\dag(t)]^2} =
  \overline{\mathrm{tr}\,[U(t)]^2\,\mathrm{tr}\,U^\dag(t)\,\mathrm{tr}\,U^\dag(t)}=\mathcal{K}_{2,11}(t)
  \\
  e&\equiv\overline{\mathrm{tr}\,[U(t)]^2 \,\mathrm{tr}\,{[U(t)U^T(t)]}}=\overline{\mathrm{tr}\,\left[U(t)U^T(t)\right]\,\mathrm{tr}\,[U^\dag(t)]^2} \\
  f&\equiv  \overline{\mathrm{tr}\,[U(t)]^2 [U^*(t)]^2}
  . 
\end{align}
In these equations, we used the invariance of $U\rightarrow U^\dag$ or $U^*$ to equate equivalent terms.

Treating the above quantities in a similar fashion to $a$, $b$, and $c$, we find that we also have to include the correlators
\begin{align}
  g&=\overline{\mathrm{tr}\,U(t)\,\mathrm{tr}\,U(t)\,\mathrm{tr}\,U^\dag(t)\,\mathrm{tr}\,U^\dag(t)} ,\\
  h&=\overline{\mathrm{tr}\,U(t)\,\mathrm{tr}\,U(t)\,\mathrm{tr}\,\left[U^\dag(t)U^*(t)\right]} ,\\
  i&=\overline{\mathrm{tr}\,U(t) \,\mathrm{tr}\,\left[U(t)U^*(t)U^*(t)\right]} ,\\
  j&=\overline{\mathrm{tr}\,\left[U(t)U^T(t)\right]\, \mathrm{tr}\,\left[U^\dag(t)U^*(t)\right]} ,\\
  l&=\overline{\mathrm{tr}\,\left[U(t)U^*(t)\right]\, \mathrm{tr}\,\left[U(t)U^*(t)\right]} ,\\
  m&=\overline{\mathrm{tr}\,U(t)\, \mathrm{tr}\,U^\dag(t)\,\mathrm{tr}\,\left[U(t)U^*(t)\right]} ,\\
  n&=\overline{\mathrm{tr}\,\left[U(t)U^*(t)U(t)U^*(t)\right]}, \\
  o&=\overline{\mathrm{tr}\, \left[U(t)U^T(t)U^\dag(t)U^*(t)\right]}. 
\end{align}

Working through the construction for all these quantities, we arrive at the system of linear differential equations
\begin{align}
  \dot{a}&=\kappa \left(\nu \left(N + b\right) -a\right) ,\\
  \dot{b}&=\kappa \left(\nu\left( a + N\right) -b\right) ,\\
  \dot{c}&= \kappa \left(\nu\left(-2(d+e) +4(N +f)\right) -2c\right) ,\\
  \dot{d}&= \kappa \left(\nu \left(-c-e-g-h+4(a+i)\right) -2d\right) ,\\
  \dot{e}&= \kappa \left(\nu \left(8b-h-j\right) -3e\right) ,\\
  \dot{f}&= \kappa \left(\nu \left(c+l+2N\right) -2f\right) ,\\
  \dot{g}&=\kappa \left(\nu \left(-2(d+h)+4(Na+m)\right) -2g\right) ,\\
  \dot{h}&=\kappa \left(\nu \left(8a-e-j\right) -3h\right) ,\\
  \dot{i}&=\kappa \left(\nu \left(a+b+d+n\right) -2i\right) ,\\
  \dot{j}&=\kappa \left(8N\nu-4j\right) \hspace{6cm} \mathrm{(closes)} ,\\
  \dot{l}&=\kappa \left(\nu(2f+2m-2n-2o+2N(b+1)) -2l\right) ,\\
  \dot{m}&=\kappa \left(\nu(g+l+N(a+b)) -2m\right) ,\\
  \dot{n}&=\kappa \left(\nu(-2l-2o+4(i+b)) -2n\right) ,\\
  \dot{o}&=\kappa \left(\nu(-j-o+4(a+b)) -3o\right) \hspace{2cm} \mathrm{(closes\,\,indirectly)}. 
\end{align}
These are then solved with their initial values corresponding to the initial condition $U(t=0)=\openone$. 
This gives, from the expressions for $c$ and $d$, the second-order instantaneous SFF
\begin{align}
  K_2(t)=&\,2+\frac{1}{6} (N-3) N (N+1) (N+2) \cosh \left(4\nu\kappa t\right)e^{-2 \kappa t}+(N-1) \left((N+2) e^{-\nu\kappa t}-N e^{\nu\kappa t}\right)e^{-3 \kappa t}+e^{-4 \kappa t}
   \nonumber \\ &-\frac{1}{12} (N-1) \left[\left(N^3+7 N^2-12 N-24\right) e^{2 \nu\kappa t}+\left(N^3-5 N^2-12 N+12\right) e^{-2 \nu\kappa t}\right]e^{-2 
 \kappa t}
 \label{eq:k22goeresult}
\end{align}
and the correlator
\begin{align}
  \mathcal{K}_{11,2}(t)=&
-\frac{1}{6} (N-3) N (N+1) (N+2) \sinh (4\nu\kappa t)e^{-2 \kappa t}
+(N-1) (N+2) e^{-(3+\nu)\kappa t}
+e^{-4 \kappa t}
    \nonumber \\ &
    +
\frac{1}{12} (N-1) \left[(N-2) \left(N^2+9 N+6\right) e^{-2 \nu\kappa t}-N \left(N^2-5 N-12\right) e^{2 \nu\kappa t}\right]e^{-2 \kappa t}
\label{eq:k112goe}
\end{align}
that appears in the expression \eqref{eq:otocresult} for the ensemble-averaged OTOC.

We note that the long-time asymptotics of these expressions still agree with the CUE.
Furthermore, results for the GSE follow from the replacement rules given in Section~\ref{sec:goegse}.

\end{widetext}

\section{Results from the theory of orthogonal polynomials}
\label{sec:orthog}

We report here the exact results for the instantaneous spectral statistics and the truncated two-point correlation function for Brownian motion over the group $\mathrm{U}(N)$ as defined by Eqs.~\eqref{eq:stochastic} and \eqref{eq:u}. 
The most recent results have been obtained from the theory of orthogonal polynomials in Ref. \cite{Forrester2024Dip-ramp-plateauUN}. 
It is worth noting that the earliest results on Brownian motion over the unitary group focused on the distributions of the group elements \cite{Ito1950BrownianGroup, Yosida1952OnSpace, Hunt1956Semi-groupsGroups}, while the seminal work from Dyson \cite{Dyson1962AMatrix} addresses the Brownian motion of the eigenvalues, which is of interest here. 

Consider the Brownian motion defined by Eqs.~\eqref{eq:stochastic} and \eqref{eq:u} in an $N$-dimensional Hilbert space.
For a unitary matrix $U(t)$ at generic time $t$, with eigenvalues $\{e^{ix_m(t)}\}_{j=1}^{N}$, $-\pi < x_m(t) \leq \pi$,
let us define the eigenvalue density $\rho_{(1),N}(x;t)$, the two-point correlation function $\rho_{2,(N)}(x,y;t)$ and the truncated two-point correlation function $\rho^T_{2,(N)}(x,y;t)$
 as
 \begin{eqnarray}
 \label{eq:spectral-corr}
  \rho_{(1),N}(x;t) &\equiv& \overline{\sum_{m=1}^N\delta (x-x_m(t))}, \nonumber \\
 \rho_{2,(N)}(x,y;t) &\equiv& \overline{\sum_{m\neq n=1}^N \delta (x-x_m(t)) \delta (y-x_n(t))}, \nonumber \\
   \rho^T_{2,(N)}(x,y;t) &\equiv& \rho_{2,(N)}(x,y;t) - \rho_{(1),N}(x;t) \rho_{(1),N}(t;t).
   \nonumber \\
\end{eqnarray}
The averages of the unitary evolution are related to the Fourier components of the spectral density and the spectral two-point function. 
Specifically, introducing the moments of the eigenvalues as \cite{Forrester2024Dip-ramp-plateauUN}
\begin{eqnarray}
  m_k^{(N)}(t) \equiv \frac{1}{N} \int_{-\pi}^{\pi} \rho_{(1),N}(x;t)e^{-ikx} dx,
\end{eqnarray}
one has 
\begin{equation}
  \overline{\textrm{tr}\,[U(t)]^k}= N m_k^{(N)}(t).
\end{equation}
Analogously, the instantaneous spectral form factor $K_n(t)$ introduced in Eq.~\eqref{eq:instantaneousSFF} is related to the Fourier transform of the two-point spectral function via 
\begin{equation}
\label{relation-SFF}
K_k(t)= S_N(k;t) + N^2 \vert 
 m_k^{(N)} (t) \vert^2,
\end{equation}
where 
\begin{equation}
  \label{eq:forrester-SFF}
  S_N(k;t) \equiv 
N + \int_{-\pi}^\pi \, dx \int_{-\pi}^\pi \, dy e^{ik (x-y)} \rho^T_{(2),N}(x,y;t).
\end{equation}

Therefore, the SFF $K_n(t)$ is determined by the connected part $S_N(k;t)$ of the SFF and the spectral density moments $m_k^{(N)}(t)$; see also Eq.~\eqref{eq:connectedsff}.

Exact results for the spectral density moments have been obtained in \cite{Onofri1981SUNAction, Andrews1984LatticeFunctions} and recently reproduced using the cyclic P\`olya ensemble structure \cite{Forrester2024Dip-ramp-plateauUN}. They can be expressed as 
\begin{equation}
  m_k^{(N)}(t)= q^{k (N+k-1)} {}_2F_1 (1-N,1-k;2;1-q^{-2k}),
\end{equation}
where $q=e^{-t/2N}$ and ${}_2F_1(a,b;c;z)$ is the Gaussian hypergeometric function.
The exact result for the truncated SFF obtained in Ref. \cite{Forrester2024Dip-ramp-plateauUN} reads as
\begin{widetext}
\begin{multline}
  S_N(k;t)= \min(N,k)-q^{2k^2+2k(N-1)} \sum_{j=0}^{k-1}\sum_{l=0}^{k-1} q^{-2k(j+l)}(-1)^{j+l} \times \\ 
   \frac{\Gamma(N+k-j)\Gamma(N+k-l)}{\Gamma(N-j)\Gamma(N+j)\Gamma(l+1)\Gamma(N-l)\Gamma(k-j)\Gamma(k-l)\Gamma(j+l-N-k+1)^2},
\end{multline}
which also admits the equivalent integral form
\begin{equation}
  S_N(k;t)=\min(k,N)-q^{2k^2+2k(N-1)}(kN)^2 \int_0^\infty s\,e^{-s(N+k-1)}\left( {}_2F_{1}(1-N,1-k;-(k-1+N);q^{-2k}e^s) \right)^2 \, ds.
\end{equation}
The latter expression can be used to obtain the large-$N$ asymptotic expansion for the connected part of the SFF at fixed $k$, which gives
\begin{equation}\label{eq:forrester-large}
  \lim_{N\to \infty} S_N(k;t) =k-e^{-tk} \int_0^{\infty} s e^{-s} \left( 
L_{k-1}^{(1)}(kt+s) \right)^2 \, ds,
\end{equation}
where $L_{s}^{(a)}(z)$ are the Laguerre polynomials.
Analogously, one can establish the large-$N$ expansion of the moments $m_k^{(N)}=A_n/N$, whose leading order follows from Eq.~\eqref{eq:unitaryleadinga}.
Carried out to order $O(N^0)$, these expansions combine to agree with the asymptotics for the SFF obtained from the hierarchy equations, as given in Eq.~\eqref{eq:asymptotics1}.
Note that the hierarchy equations also yield exact and asymptotic expansions for the additional correlations $\mathcal{K}_{pq\dots,rs\dots}(t)$ defined in Eq.~\eqref{eq:splittracecue}.

\end{widetext}

\section{Stationary solutions in the unitary hierarchy}
\label{app:ergodicother}

As discussed in the main text, in the long-time limit the instantaneous SFF converges, in all three symmetry classes, to the RMT prediction \eqref{eq:sffcue} for an ergodic system with unitary invariance, where the unitary time-evolution operators $U(t)$ are taken from the CUE. 

This agreement in the ergodic limit also extends to the long-time behavior of the other quantities $\mathcal{K}_{\{\cdot\},\{\cdot\}}(t)$ appearing in the hierarchy, for which the CUE results have been established as well \cite{Haake1996SecularMatrices, Diaconis2001LinearMatrices, Diaconis2004RandomPolynomials}. 
These results can be expressed compactly as 
\begin{align} \label{eq:rmtavg}
\overline{\prod_{j=1}^k \left(\mathrm{tr}\, \left[U\right]^j\right)^{a_j}\, \left(\mathrm{tr}\, \left[U^\dag\right]^j\right)^{b_j}} &= \delta_{ab}\,\prod_{j=1}^k j^{a_j}{a_j}!, 
\end{align}
where $a=(a_1,...a_k)$ and $b=(b_1,...b_k)$ are two integer sequences, while the overline denotes the averaging of $U$ in the CUE.
These expressions are consistent with the stationary solutions of the spectral hierarchy, which is determined by setting all time derivatives to zero. 
In particular, we recover that only balanced quantities \eqref{eq:define} of the form $\{\cdot\}_\mathrm{LHS} = \{\cdot\}_\mathrm{RHS}$ tend to nonzero stationary values. These values furthermore correspond to Eq.~\eqref{eq:rmtavg} when we identify $a_j$ with the number of times an integer $j$ appears in the index sequence $\{\cdot\}_\mathrm{LHS}$.

\begin{figure}[b]
  \centering
  \includegraphics[width=\linewidth]{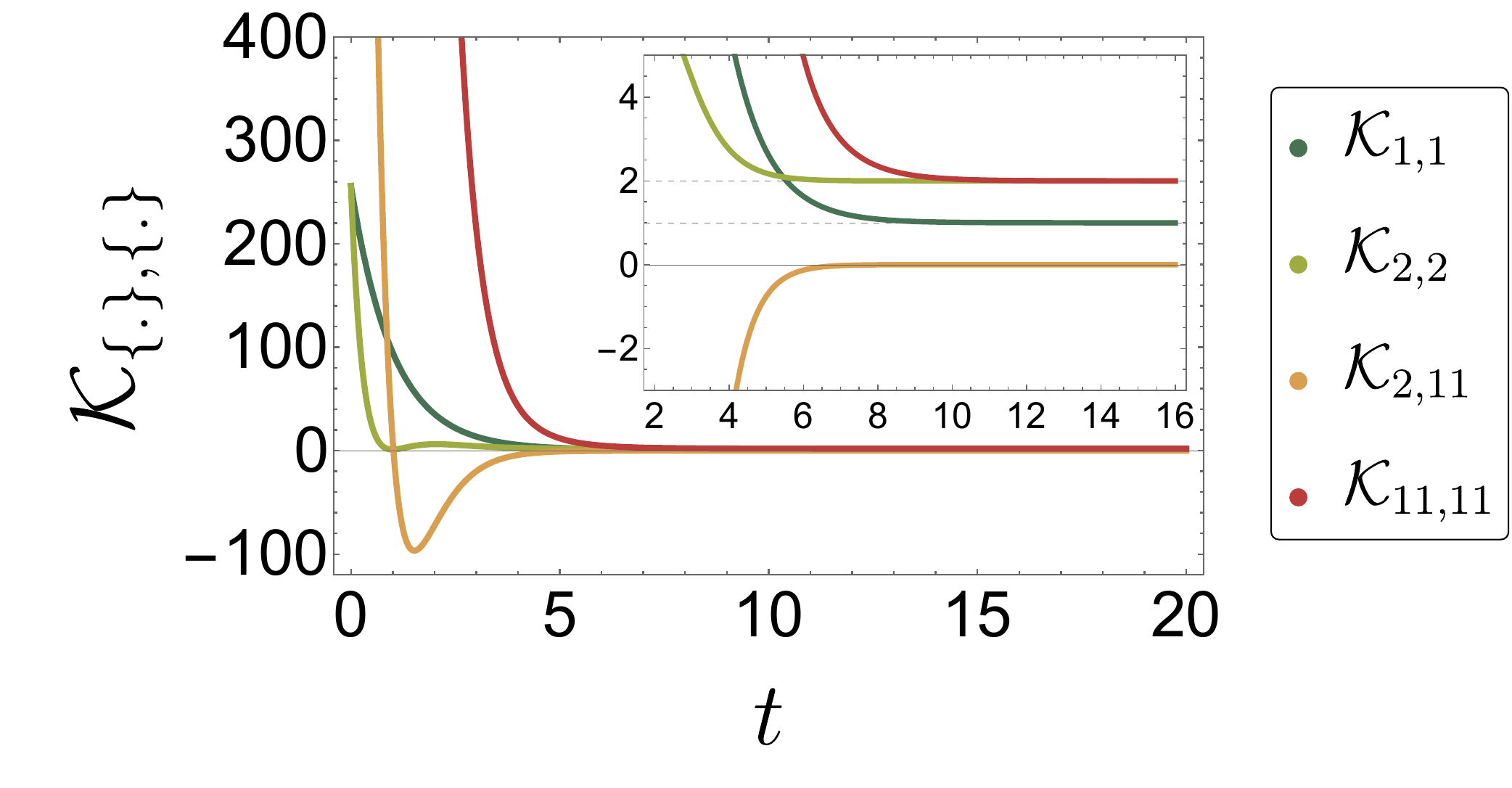}
  \caption{Analytical solutions of all correlators appearing in unitary spectral hierarchy up to level $n=2$, for system size $N=16$. The SFF 
  $\mathcal{K}_{n,n}(t)$ (light and dark greens) tends to $n$ at longer times, as indicated by the grey dashed lines. The correlator $\mathcal{K}_{11,11}(t)$ (red) is the only other quantity at this level that tends to a nonzero stationary value, given by 2. The correlator $\mathcal{K}_{2,11}(t)$ (orange), and equivalently $\mathcal{K}_{11,2}(t)$, decays to 0. All of these long-time asymptotics agree with the general CUE result \eqref{eq:rmtavg}. 
  }
  \label{fig:n=2}
\end{figure}

This behavior is further illustrated in
Figure~\ref{fig:n=2}, which depicts the analytical solutions of all quantities in the unitary spectral hierarchy to level $n=2$. These solutions are obtained from Eqs.~\eqref{eq:system2} paired with the initial conditions \eqref{eq:system2initial}, which are evaluated for $N=16$. 
All balanced terms tend to integer values: the first- and second-order SFF $\mathcal{K}_{1,1}(t)$ and $\mathcal{K}_{2,2}(t)$, tend to $1$ and $2$, as already discussed above, while $\mathcal{K}_{11,11}(t)$ tends to $2$, as dictated by Eq.~\eqref{eq:rmtavg}. 
Both $\mathcal{K}_{11,2}(t)$ and $\mathcal{K}_{2,11}(t)$ are unbalanced in their indices, thus tend to zero, and coincide by way of symmetry, thus are overlaid.

\end{document}